\documentclass[aps,prb,twocolumn,groupedaddress,floatfix,superscriptaddress]{revtex4-2}

\usepackage{epsfig}
\usepackage{graphicx}
\usepackage[normalem]{ulem}
\usepackage{mathtools}
\usepackage{amsmath,amssymb}
\usepackage{physics}
\usepackage{hyperref}
\hypersetup{colorlinks=true,citecolor={blue},linkcolor={blue},urlcolor={blue}}
\usepackage[dvipsnames,usenames]{xcolor}

\newcommand{\pcsadd}{Center for Theoretical Physics of Complex Systems, Institute for Basic Science(IBS), Daejeon, Korea, 34126}
\newcommand{\ustadd}{Basic Science Program, Korea University of Science and Technology (UST), Daejeon 34113, Republic of Korea}
\newcommand{\cqtadd}{Centre for Quantum Technologies, National University of Singapore, 3 Science Drive 2, 117543 Singapore}
\newcommand{\dpnusadd}{Department of Physics, National University of Singapore, 2 Science Drive 3, 117551 Singapore}

\newcommand{\lmax}{L_\mathrm{max}}

\begin{document}

\title{Superconductivity with Wannier-Stark Flat Bands} 

\author{Si Min Chan}
    \affiliation{\cqtadd}
    \affiliation{\dpnusadd} 

\author{Alexei Andreanov}
    \affiliation{\pcsadd}
    \affiliation{\ustadd}

\author{Sergej Flach}
    \affiliation{\pcsadd}
    \affiliation{\ustadd}
 
\author{G. George Batrouni}
    \affiliation{\cqtadd} 
    \affiliation{\dpnusadd} 
    \affiliation{Universit\'e C\^ote d'Azur, CNRS, Institut de Physique de Nice (INPHYNI), 06103 Nice, France} 

\begin{abstract}
    We investigate superconducting transport in the DC field induced Wannier-Stark flat bands in the presence of interactions. 
    Flat bands offer the possibility of unconventional high temperature superconductivity, where the superfluid weight, \(D_s\), is enhanced by the density overlap of the localized states.
    However, construction of flat bands typically requires very precise tuning of Hamiltonian parameters.
    To overcome this difficulty, we propose a feasible alternative to realize flat bands by applying a DC field in a commensurate lattice direction.
    We systematically characterize the superconducting behavior on these flat bands by studying the effect of the DC field and attractive Hubbard interaction strengths on the wavefunction, correlation length, pairing order parameter and the superfluid weight \(D_s\). 
    Our main result is that the superfluid weight is dramatically enhanced at an optimal value of the interaction strength and weak DC fields.
\end{abstract}

\maketitle

\section{Introduction}

Topologically nontrivial flat bands have proven to be promising in the search for high temperature
superconductivity~\cite{cao2018unconventional,chan2022pairing,chan2022designer,
  xie2020topology, huhtinen2022revisiting, tovmasyan2016effective,
  peotta2015superfluidity, peri2021fragile,julku2016geometric,
  pyykkonen2023suppression, torma2022superconductivity,iglovikov2014superconducting}.
The merit of superconductivity on isolated flat bands is two-fold:
(a) due to quenched kinetic energy any finite interaction is relevant, resulting in a strongly correlated system, where superconducting transport arises from the correlated hopping of the fermions;
and (b) both the critical temperature (\(T_c\)) and superfluid weight (\(D_s\)) depend linearly on the attractive interaction (\(U\)), for weak interaction, thus is exponentially enhanced compared to dispersive bands with weak attraction between electrons~\cite{julku2016geometric,
  peotta2015superfluidity,heikkila2016flat, kopnin2011high,
  peri2021fragile, heikkila2011flat}.
In these systems, flat band superconductivity at weak interaction was shown to be dominated by a single isolated flat band~\cite{peotta2015superfluidity, huhtinen2022revisiting, chan2022pairing, tovmasyan2016effective}.
The superconducting properties of the many-body system, in presence of a weak attraction, are largely dependent on both the
wavefunction and the band topology in the non-interacting  limit~\cite{peotta2015superfluidity,torma2022superconductivity,peri2021fragile,julku2016geometric,tovmasyan2016effective}.

However, there is a caveat: in isolated flat bands with compact localized states (CLS), the CLS must span more than one unit cell in order for the linear dependence \(D_s\propto U\) and \(T_c \propto U\) to emerge~\cite{chan2022designer,peotta2015superfluidity,torma2022superconductivity, verma2021optical}.
To optimize \(D_s\) further, the system parameters must be chosen such that the overlap between neighboring CLS is maximized, resulting in the maximization of the density overlap of the Wannier functions~\cite{chan2022designer, tovmasyan2016effective}.

The construction of flat bands typically requires subtle fine-tuning of Hamiltonian parameters.
To design flat bands, a specific set of rules have to be followed and hopping potentials are fixed with precise values chosen to ensure flatness of the  band~\cite{maimaiti2019universal,maimaiti2021flat,cualuguaru2022general,kim2023general,hwang2021general}.
While these systems have been experimentally realized in superconducting wire networks, atomic lattices and optical lattices, flat bands with longer-range hopping, e.g. beyond nearest neighbours, 
which provide control over the CLS, become difficult to fabricate~\cite{leykam2018artificial, huda2020designer,vidal1998aharonov, abilio1999magnetic, jo2012ultracold, hyrkas2013many, vicencio2015observation}.
Flat band superconductivity was observed experimentally in twisted bilayer graphene, but the bilayers have to be stacked precisely at a `magic' angle at which band flatness is maximized (though not being perfectly flat)~\cite{cao2018unconventional}.
Is there a more accessible way to generate flat bands, \emph{and} control the spread of particle wavefunction?

When a static DC field is oriented in a lattice commensurate direction of a \(D\geq2\) Bravais lattice, it was shown that flat bands of dimension \(D-1\) are formed in the direction perpendicular to the DC field, and the emergence of flat bands is independent of the field strength~\cite{mallick2021wannier}.
These Wannier-Stark flat bands do not have CLS, only superexponentially localized wavefunctions which depend on the field strength.
We propose that this method of generating flat bands comes with two advantages for superconductivity:
(a) no fine-tuning is required and (b) there is no CLS which, therefore, increases the density overlap of the wavefunctions and potentially enhances supercondutcivity.
In practice, one can, for example, emulate this in ultracold atomic gas experiments, by tilting the system at an angle and the gravitational field acts in the chosen commensurate direction of the optical  lattice~\cite{anderson1998macroscopic,ivanov2008coherent,roati2004atom,sapienza2003optical}.

Previous studies have shown extensively that the multi-band mean-field theory describes the superconducting properties of flat band systems remarkably well in comparison with density matrix renormalization group (DMRG) computations and determinant quantum Monte-Carlo  (DQMC)~\cite{chan2022pairing,chan2022designer,huhtinen2022revisiting,tovmasyan2016effective,julku2016geometric,iglovikov2014superconducting}.
Thus in this study, we investigate interacting fermions in the Wannier-Stark flat band systems using the full multi-band mean-field theory~\cite{chan2022pairing,chan2022designer}.
In particular, we show the dependence of the superfluid weight, pairing order parameters, lattice filling and correlation functions on both the DC field and interaction strengths.

Our main result is that the flat band enhancement of superconductivity persists in Wannier-Stark systems despite the trivial topology of the underlying lattice.
Consequently, superconductivity can be readily improved in Bravais lattices by applying a static field in a commensurate direction, which induces Wannier-Stark flat bands.
These flat bands act as superconducting transport channels with the additional advantage of a tuneable spread of the wavefunction and band gaps.
At weak Hubbard attraction and sufficiently strong DC field, the pairing order parameter is enhanced, while the superfluid weight is large due to the transport being supported by multiple flat bands, growing linearly with the interaction.
Moreover, for sufficiently large DC field strength, the correlation length is smaller than a lattice spacing, typical for flat bands, indicating bound Cooper pairs are very small, essentially on a single site.

The paper is organized as follows.
We introduce the Hubbard Hamiltonian on the Wannier-Stark lattice in 2D and 3D in Section~\ref{sec:model}.
Additionally, we briefly outline how to compute the superconducting quantities.
Thereafter, in Section~\ref{sec:Results}, we characterize the superconducting properties of Wannier-Stark flat bands and develop an understanding of its dependence on the DC field and Hubbard interaction strengths.
Finally, we conclude with the main takeaways from this study and future outlook that we propose in Section~\ref{sec:conclusions}.

\section{Model and Methods}
\label{sec:model}

We begin by considering the 2D square lattice with a DC field \(E\mathbf{e_z}\) oriented in a commensurate direction of the lattice, and the on-site attractive Hubbard interaction term.
The Hamiltonian can be expressed with the single particle and interaction components, as
\begin{equation}
    H-\mu N= H_S+H_I
\end{equation}
where
\begin{equation}
\begin{aligned}
    H_S=&-t\displaystyle\sum_{x,y,\sigma}\left(c_{x,y,\sigma}^\dagger c_{x+1,y,\sigma}+c_{x,y,\sigma}^\dagger c_{x,y+1,\sigma}+ \mathrm{h.c.}\right) \\
    &+\displaystyle\sum_{x,y,\sigma}(E\mathbf{e_z\cdot r}-\mu) c_{x,y,\sigma}^\dagger c_{x,y,\sigma} \\  
    H_I=&-U\displaystyle\sum_{x,y} c_{x,y,\uparrow}^\dagger c_{x,y,\uparrow}c_{x,y,\downarrow}^\dagger c_{x,y,\downarrow} 
\end{aligned}
\end{equation}
Here, \(\mathbf{r} = xa\mathbf{e_x}+ya\mathbf{e_y}\), with \(x,y \in \mathbb{Z}\) and \(a\) is the lattice constant.
\(c_{x,y,\sigma}^\dagger (c_{x,y,\sigma})\) creates (annihilates) fermions of spin \(\sigma \in (\uparrow, \downarrow)\) on site \((x,y)\).
\(t\) is the hopping parameter, \(\mu\) is the chemical potential which controls the filling of the lattice and \(U>0\) is the
Hubbard interaction strength.
We define the orthogonal vectors \(\mathbf{e_z}\) and \(\mathbf{e_w}\) which are respectively parallel and perpendicular to the DC field direction.
For the commensurate DC field direction \(\mathbf{e_w}\) is parallel to some lattice vector.
For specific calculations, we set \(\mathbf{e_z} = (\mathbf{e_x}- \mathbf{e_y})/\sqrt{2}\) and \(\mathbf{e_w} =  (\mathbf{e_x}+\mathbf{e_y})/\sqrt{2}\) in this study, where lattice sites can then be labelled with \(z,w \in \mathbb{Z}\).

\begin{figure}
    \includegraphics[width=7cm]{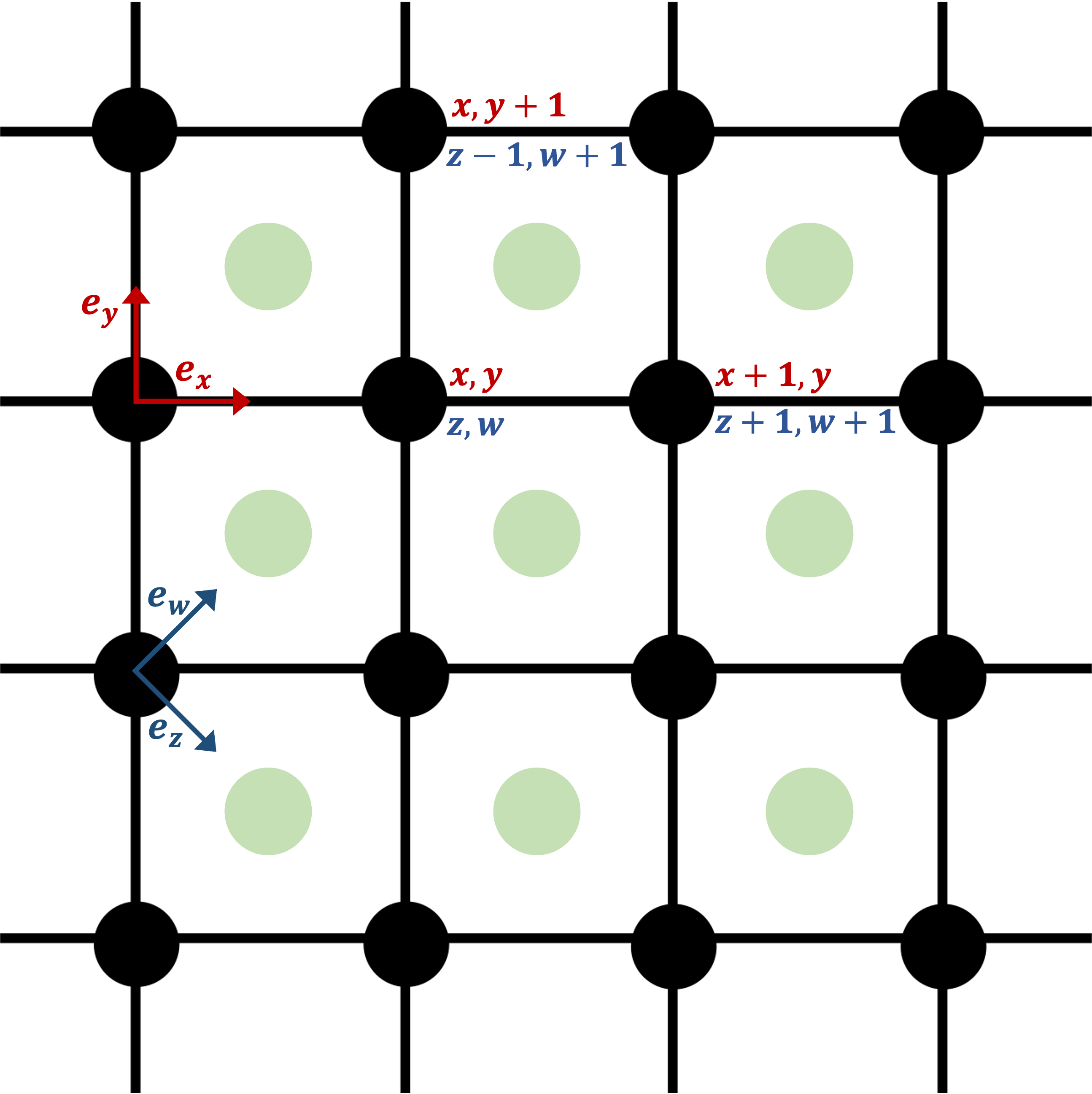}
    \caption{
        (Color online) Square lattice with DC field direction \(\mathbf{e_z}\) indicated.
        Original and new coordinates of the square lattice are shown as \(x,y\) and \(z,w\).
        In the \(z,w\) basis, both coordinates have to be simultaneously changed (\(\pm 1\)) to describe hopping in the original basis.
    }
    \label{fig:squarelattice}
\end{figure}

The single particle Hamiltonian in the \(z, w\) basis is
\begin{equation}
    \label{eq:Hs_2D}
    \begin{aligned}
        H_S=&-\frac{t}{2}\displaystyle\sum_{z,w,\sigma} \left(c_{z,w,\sigma}^\dagger c_{z+1,w+1,\sigma}+c_{z,w,\sigma}^\dagger
        c_{z-1,w+1,\sigma}+\mathrm{h.c.}\right) \\
        &+\frac{1}{2}\displaystyle\sum_{z,w,\sigma}(Fz-\mu) c_{z,w,\sigma}^\dagger c_{z,w,\sigma}
    \end{aligned}
\end{equation}
with \(F = \frac{aE}{\sqrt{2}}\).

\begin{figure}
    \includegraphics[width=7cm]{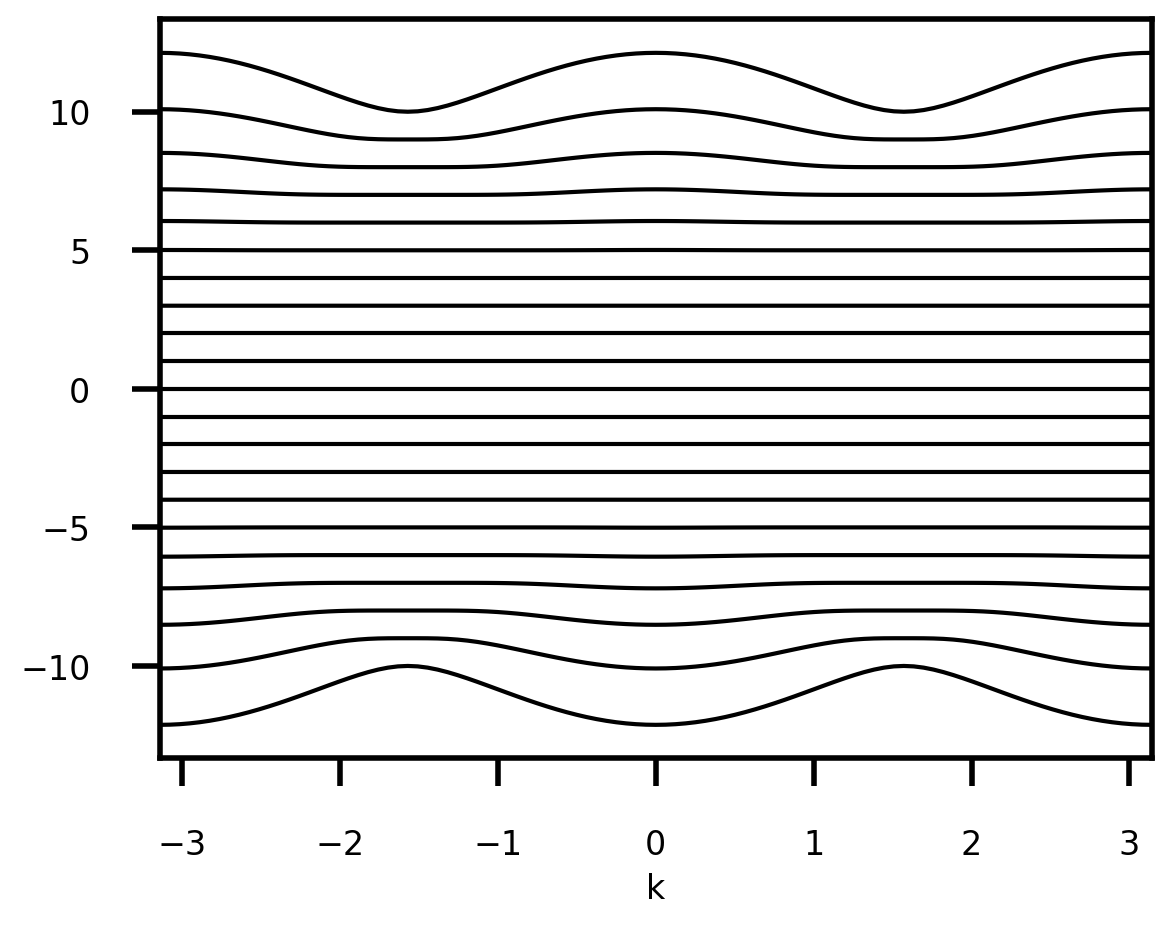} \\
    \includegraphics[width=7cm]{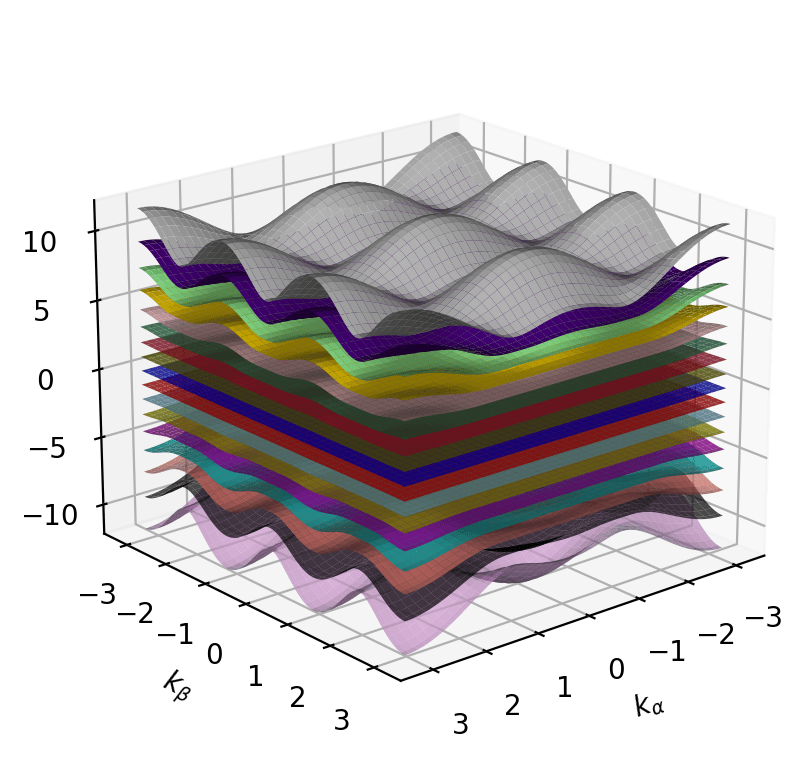}
    \caption{ 
        (Color online) \(D-1\) Wannier-Stark flat bands with finite a \(\lmax\).
        Bands acquire finite dispersion at the edges of the spectrum due to finite size effects.
        \textbf{Top}: the 2D square lattice with \(F=1\) and \textbf{Bottom}: the 3D cubic lattice with \(F=1\).
    }
    \label{fig:WannierStarkBands}
\end{figure}

In the \(z,w\) basis, we apply the mean-field decoupling on the Hubbard interaction term (Appendix \ref{appendix:multibandMF}),
recognizing that the mean-field parameters \(\Delta^z/U = \langle c_{z,w,\uparrow}c_{z,w,\downarrow}\rangle\) and \(\rho^z_\sigma=\langle c_{z,w,\sigma}^\dagger c_{z,w,\sigma}\rangle\) should depend on \(z\),
\begin{equation}
    \label{eq:Hint_2D}
    \begin{aligned}
        H_I=&-\frac{U}{2}\displaystyle\sum_{z,w} c_{z,w,\uparrow}^\dagger c_{z,w,\uparrow}c_{z,w,\downarrow}^\dagger c_{z,w,\downarrow} \\
        \widetilde{H}_I=& - \frac{1}{2}\displaystyle\sum_{z,w}\left(\Delta^z c_{z,w,\downarrow}^\dagger c_{z,w,\uparrow}^\dagger + \Delta^{z*} c_{z,w,\uparrow}c_{z,w,\downarrow}\right) \\ 
        &-\frac{U}{2}\displaystyle\sum_{z,w}\left(\rho^z_\uparrow c_{z,w,\downarrow}^\dagger c_{z,w,\downarrow} + \rho^z_\downarrow c_{z,w,\uparrow}^\dagger c_{z,w,\uparrow} \right) \\
        &+\frac{1}{2}L_w\displaystyle\sum_{z} \left(U\rho^z_{\uparrow}\rho^z_{\downarrow} + \abs{\Delta_{z}}^2/U\right) 
    \end{aligned}
\end{equation}
where \(\widetilde{H}_I\) is the mean field approximation of the interaction term, and \(L_w\) is the number of sites in the \(w\) direction.
Here, \(\Delta^z/U\) and \(\rho^z_\sigma\) are the pairing order parameter and filling on sites with equal values of \(z\).  
The factor \(\frac{1}{2}\) in Eqs.~\eqref{eq:Hs_2D} and~\eqref{eq:Hint_2D} accounts for double counting and arises when we sum over \(z\) and \(w\) due to the two sublattices as shown in Fig~\ref{fig:squarelattice}.
While the translational symmetry in the \(\mathbf{e_z}\) direction is broken due to the DC field, the system remains translationally invariant in the \(\mathbf{e_w}\) direction.
Under Fourier transformation in \(w\), we can express the Hamiltonian as
\begin{widetext}
\begin{equation}
\begin{aligned}
    2(H-\mu N) &= -t\displaystyle\sum_{z,k,\sigma}\left(c_{z,k,\sigma}^\dagger c_{z+1,k,\sigma}e^{ik}+c_{z,k,\sigma}^\dagger c_{z-1,k,\sigma}e^{ik} + 
    h.c.\right)+\displaystyle\sum_{z,k,\sigma}(Fz-\mu) c_{z,k,\sigma}^\dagger c_{z,k,\sigma} \\ 
    &- \displaystyle\sum_{z,k}\left(\Delta c_{z,-k,\downarrow}^\dagger c_{z,k,\uparrow}^\dagger + \Delta^*
    c_{z,k,\uparrow}c_{z,-k,\downarrow}\right) - U\displaystyle\sum_{z,k}\left(\rho^z_\uparrow
    c_{z,k,\downarrow}^\dagger c_{z,k,\downarrow} + \rho^z_\downarrow c_{z,k,\uparrow}^\dagger c_{z,k,\uparrow} \right) \\ 
    &=\displaystyle\sum_{k}\Psi_{k}^\dagger \mathcal{M}(k)\Psi_{k}+L_w\displaystyle\sum_{z}
    \left(U\rho^z_{\uparrow}\rho^z_{\downarrow} +\abs{\Delta_{z}}^2/U\right) +
    L_w\displaystyle\sum_z(Fz-\mu-U\rho^z_\uparrow) \\ 
    \mathcal{M}(k) &= 
    \begin{pmatrix}
        \mathcal{T}(k) & \mathcal{D} \\
        \mathcal{D}^* & -\mathcal{T}^\mathsf{T}(-k)
    \end{pmatrix},\\
    \mathcal{T}(k) &= 
    \begin{pmatrix} 
        &  & &  \ddots &  &  &  \\
        & ... & 0 & -2t\cos(k) & -F-\mu-U\rho^{z=-1}_\downarrow & -2t\cos(k) & 0 & ... \\
        & & ...  & 0 & -2t\cos(k) & -\mu-U\rho^{z=0}_\downarrow & -2t\cos(k) & 0 & ... &  \\
        & & & ... & 0 & -2t\cos(k) & F-\mu-U\rho^{z=1}_\downarrow & -2t\cos(k) & 0 & ... \\
        & & & &  &  &  & \ddots & 
    \end{pmatrix},\\
    \mathcal{D} &=\mathrm{diag}(\Delta^{z=-L_\mathrm{max}}, ...,
    \Delta^{z=-1}, \Delta^{z=0}, \Delta^{z=1}, ...,
    \Delta^{z=L_\mathrm{max}}), 
\end{aligned}
\end{equation}
\end{widetext}
the last term in the Hamiltonian results from anti-commutation relations, \(\Psi_{z,k}^\dagger\) is the Nambu spinor and the matrix \(\mathcal{M}(k)\) is the Bogoliubov-de Gennes Hamiltonian.

\(2\lmax+1\) is the total number of sites in the \(\mathbf{e_z}\) direction.
In the absence of interaction, we obtain \(2\lmax+1\) one-dimensional flat bands along the \(\mathbf{e_w}\) direction, each labeled by index \(z\), for any finite value of \(F\).
The wavefunctions of the flat band eigenstates are superexponentially localized, and can be expressed analytically via Bessel functions.
In the case where the DC field is oriented along the main diagonal, the extents of the wavefunction in the \(\mathbf{e_w}\) and
\(\mathbf{e_z}\) directions are the same~\cite{mallick2021wannier}.
For a particle at \((z=0,w=0)\),
\begin{equation}
    \label{eq:wavefunctions}
    \begin{aligned}
        \Psi(z=0,r_w)&=J_{-r_w}\left(-\frac{2t}{F}\right)J_{-r_w}\left(\frac{2t}{F}\right) \\ 
        \Psi(r_z,w=0)&=J_{-r_z}\left(-\frac{2t}{F}\right)J_{r_z}\left(\frac{2t}{F}\right)  
    \end{aligned}
\end{equation}
where \(r_w\) and \(r_z\) are integers indexing the number of sites away from the center, and \(J_n(x)\) are Bessel functions of the first kind and decay at least as a factorial, \(1/n!\), which is faster than exponential.
In the limit \(\lmax\to \infty, F\neq 0\), there will be an infinite number of flat bands labelled by \(z\) in the \(w\) direction.
Each flat band corresponds directly to sites with equal \(z\) coordinates, where consecutive flat bands have a gap of \(E_\mathrm{gap}=F\).
However, in real systems, and computationally, \(\lmax\) would be finite, resulting in dispersive bands near the edges \(z=\pm L_\mathrm{max}\), as shown in Fig.~\ref{fig:WannierStarkBands}.
In order to characterize accurately flat band superconductivity, these dispersive bands have to be fully occupied or empty.

Similarly, in 3D systems, one can orient the DC field along commensurate directions, and obtain flat bands in the planes perpendicular to it.
Here, we consider the cubic lattice with a DC field of strength \(E\) in the \(\mathbf{e_\gamma}\) direction, and  define the orthonormal vectors
\begin{equation}
    \mathbf{e_\alpha} = \frac{1}{\sqrt{2}}
    \begin{pmatrix} 
        -1 \\ 1 \\ 0 
    \end{pmatrix}, 
    \mathbf{e_\beta} = \frac{1}{\sqrt{6}}
    \begin{pmatrix} 
        -1 \\ -1 \\ 2 
    \end{pmatrix}, 
    \mathbf{e_\gamma} = \frac{1}{\sqrt{3}}
    \begin{pmatrix} 
        1 \\ 1 \\ 1
    \end{pmatrix}
\end{equation}
as the new basis.  
We can then express the mean field Hamiltonian in momentum space, 
\begin{widetext}
\begin{equation}
    \label{eq:3DHam}
    \begin{aligned}
        6(H-\mu
        N)=&-t\displaystyle\sum_{k_\alpha,k_\beta,\gamma,\sigma}
        c_{k_\alpha,k_\beta,\gamma,\sigma}^\dagger
        c_{k_\alpha,k_\beta,\gamma+1,\sigma}\left(2\cos(k_\alpha)e^{-ik_\beta}+e^{i2k_\beta}\right)
        + \mathrm{h.c.} \\ 
        &+\displaystyle\sum_{k_\alpha,k_\beta,\gamma,\sigma}
        \left(\frac{aE\gamma}{\sqrt{3}}-\mu\right)
        c_{k_\alpha,k_\beta,\gamma,\sigma}^\dagger 
        c_{k_\alpha,k_\beta,\gamma,\sigma}-U\displaystyle
        \sum_{k_\alpha,k_\beta,\gamma} \left(\rho_\uparrow^\gamma
        c_{k_\alpha,k_\beta,\gamma,\downarrow}^\dagger
        c_{k_\alpha,k_\beta,\gamma,\downarrow} +
        \rho_\downarrow^\gamma
        c_{k_\alpha,k_\beta,\gamma,\uparrow}^\dagger
        c_{k_\alpha,k_\beta,\gamma,\uparrow} \right) \\ 
        &-\displaystyle\sum_{k_\alpha,k_\beta,\gamma}\left(\Delta^\gamma
        c_{k_\alpha,k_\beta,\gamma,\downarrow}^\dagger
        c_{-k_\alpha,-k_\beta,\gamma,\uparrow}^\dagger +
        \Delta^{\gamma*} c_{k_\alpha,k_\beta,\gamma,\uparrow}
        c_{-k_\alpha,-k_\beta,\gamma,\downarrow}\right) +L_\alpha
        L_\beta
        \displaystyle\sum_{\gamma}\left( U\rho_\uparrow^\gamma
        \rho_\downarrow^\gamma +
        \frac{\abs{\Delta^\gamma}^2}{U}\right) 
    \end{aligned}
\end{equation}
\end{widetext}
where \(-\lmax\leq\gamma\leq \lmax, \gamma \in \mathbb{Z}\), giving rise to \(2\lmax+1\) quasi two-dimensional flat bands in the tight-binding description shown in Fig.~\ref{fig:WannierStarkBands}.
Factor \(6\) is due to the sites in the new basis that do not exist originally.
In the 3D system, the flat bands are labelled with \(\gamma\).

To characterize the superconducting state, we calculate the pairing order parameter \(\Delta^z/U\) in 2D (\(\Delta^\gamma/U\) in 3D), the superfluid weight and the single particle Green's function.
The set of mean-field parameters, \(\rho^{z(\gamma)}\) and \(\Delta^{z(\gamma)}/U\), in the presence of interaction are obtained by iteratively solving the self-consistent equations (Appendix~\ref{appendix:multibandMF}).

The superfluid weight, \(D_s\), at \(T=0\) is calculated by applying a phase twist \(\Phi\) (Appendix~\ref{appendix:Ds}) in the direction perpendicular to the DC field,
\begin{equation}
    \label{eq:Dscompare}
    \begin{aligned}
        &D_s(q1D) = \pi L_w \left. \frac{d^2E_\mathrm{GS}}{d\Phi^2}
      \right|_{\Phi=0}, \\
        &D_s(q2D) = \pi \left.\frac{d^2E_\mathrm{GS}}{d\Phi^2}
      \right|_{\Phi=0},
    \end{aligned}
\end{equation}
where the \((q1D)\) indicates the quasi one-dimensional system, and \((q2D)\) the quasi two-dimensional one. \(E_\mathrm{GS}\) is the ground state energy~\cite{hayward95,kohn64,scalapino93}.
While results are presented for both systems, we focus our analysis on the 2D (quasi-1D) Wannier-Stark lattice.

Mean-field calculations can also accurately provide information about the single-particle correlation function along and perpendicular to the field in the presence of interactions, with
\begin{equation}
    \label{eq:GF}
    \begin{aligned}
        G_\sigma(r_w) &= \langle c_{z,w,\sigma} c_{z,w+r_w,\sigma}^{\dagger}\rangle \\
        G_\sigma(r_z) &= \langle c_{z,w,\sigma} c_{z+r_z,w,\sigma}^{\dagger}\rangle
    \end{aligned}
\end{equation}
where \(r_w, r_z \in \mathbb{Z}\), and we compute these quantities for the 2D Wannier-Stark lattice to obtain insights about the spatial extent of the Cooper pairs~\cite{giamarchi2003quantum}.

In this study, we choose \(\mu=-\frac{U}{2}\) to obtain a half-filled \(z=0\) (\(\gamma=0\) in 3D) flat band, and choose sufficiently large \(\lmax\), such that all dispersive bands due to finite size effects are either completely full or empty.
All energies are measured in the units of the hopping potential \(t=1\), the lattice constant is fixed at \(a=1\) and the temperature is \(T=0\).

\section{Results and Discussion}
\label{sec:Results}

\subsection{2D Square Lattice}

We define \(F=\frac{aE}{\sqrt{2}}\) for the square lattice with the DC field of strength \(E\) along the main diagonal.
In the presence of a DC field, \(E_\mathrm{gap}=F\).
As we increase \(F\), energy gaps between neighboring diagonals become bigger and the wavefunctions are increasingly localized in real space.
With the interplay between the DC field strength and attraction, we find a strong dependence of the \(z\)-band filling and the pairing order parameter on these parameters.

\begin{figure}
    \centering
    \includegraphics[width=8.6cm]{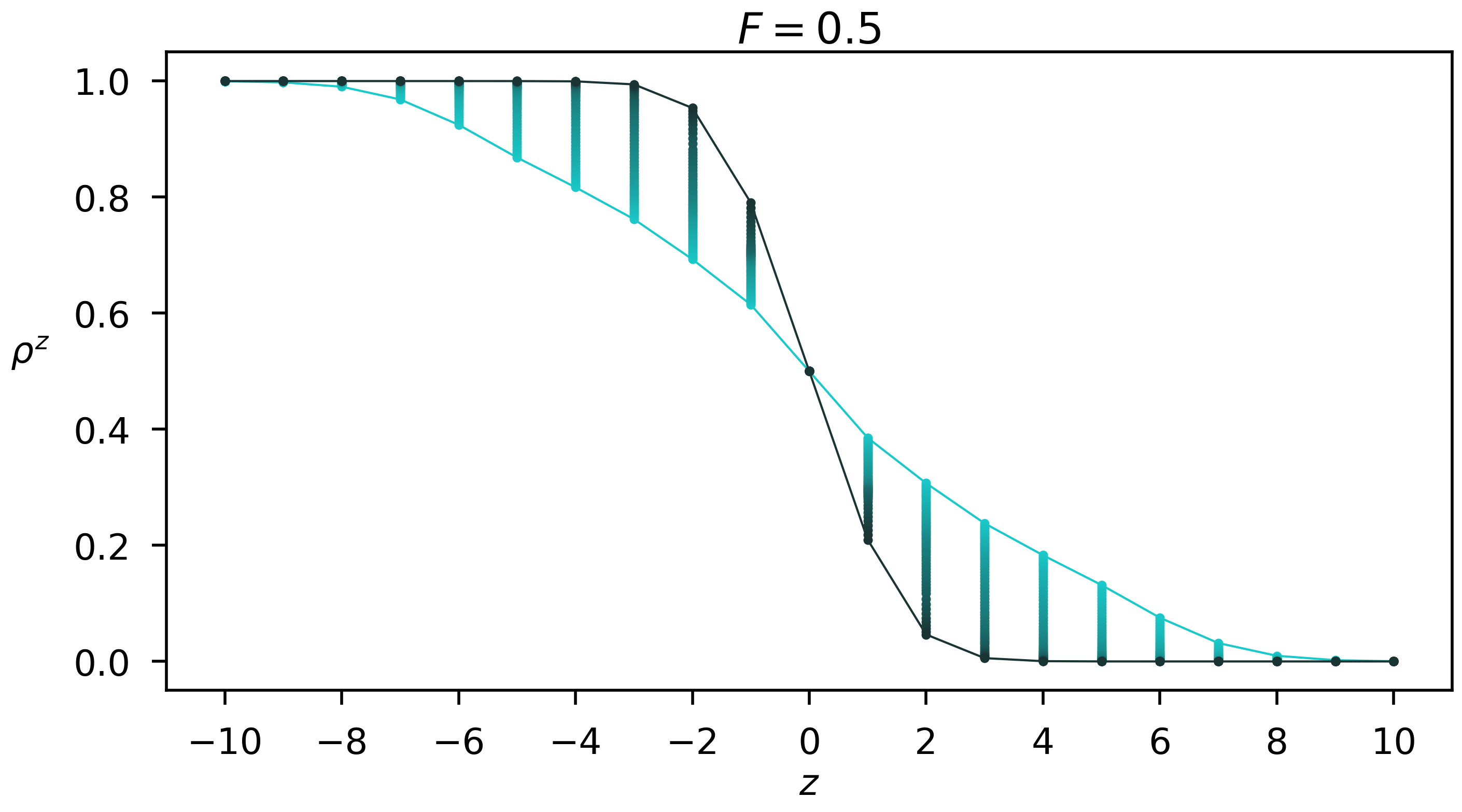} \\
    \includegraphics[width=8.6cm]{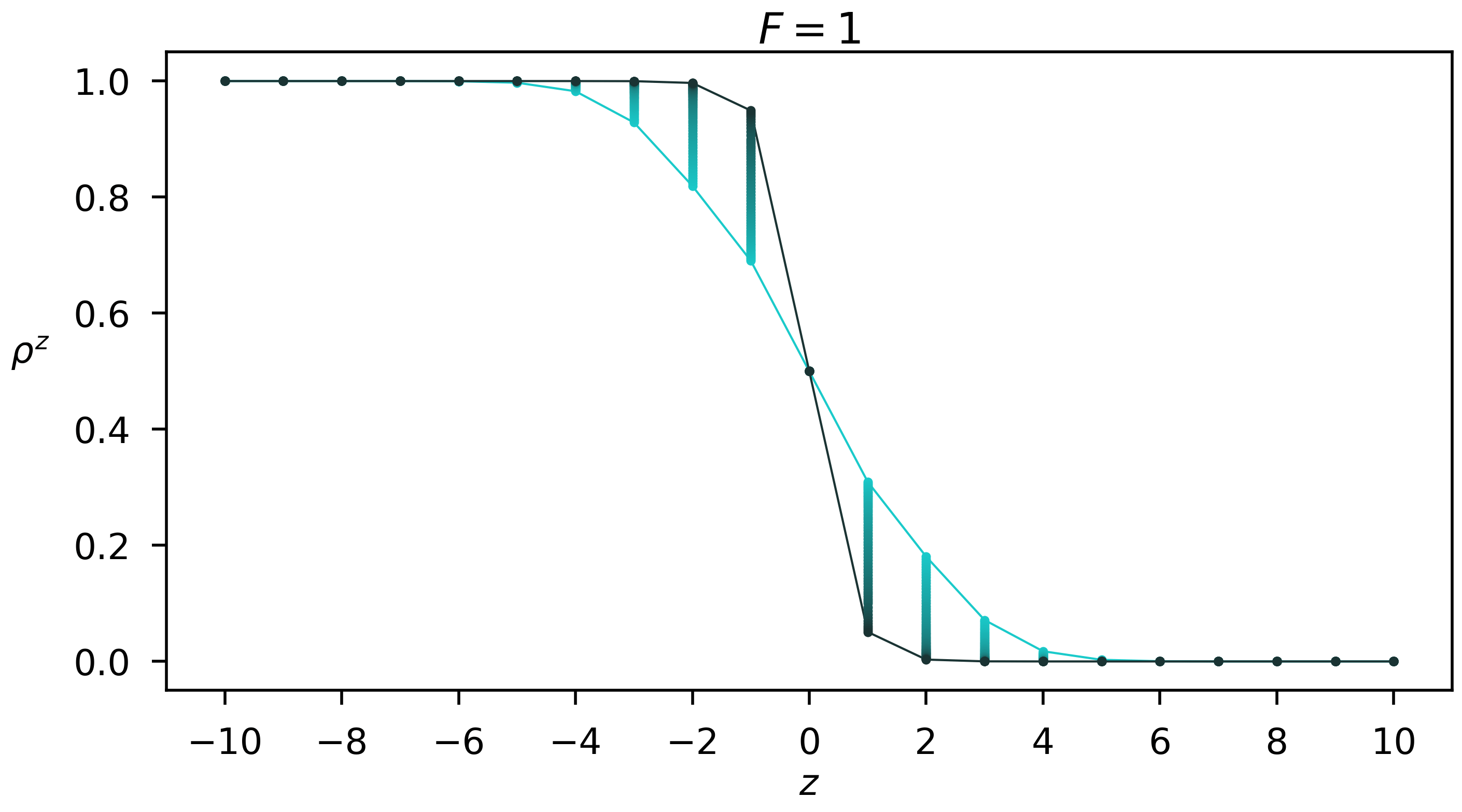} \\
    \raggedleft\includegraphics[width=7.6cm]{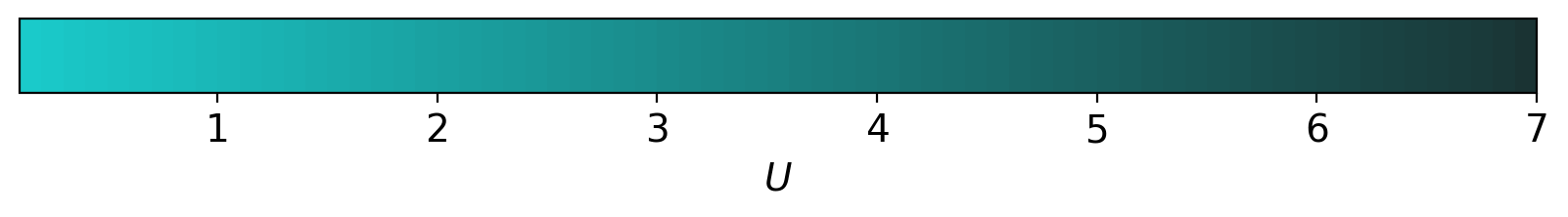}
    \caption{ 
        (Color online) Values of \(z\)-band filling, \(\rho^z\), for \(0.1\leq U\leq 7\) with \textbf{Top}: \(F = 0.5\) and \textbf{Bottom}: \(F=1\), where light (dark) color represents weaker (stronger) interaction strength.
        Solid lines are shown for \(U=0.1\) and \(U=7\).
        Since we have equal populations for the up and down spins, \(\rho_\uparrow^z = \rho_\downarrow^z=\rho^z\).
        \(z\)-bands with \(0<\rho^z<1\) are partially filled, and \(\rho^{z=0}=0.5\) for all values of \(U\).
        Due to the larger energy gap between the flat bands as \(F\) increases, the number of partially filled bands decreases.
        With stronger attractive interaction, the number of bands with partial filling also decreases.
    }
    \label{fig:dens_square}
\end{figure}

\begin{figure}
    \centering
    \includegraphics[width=8.6cm]{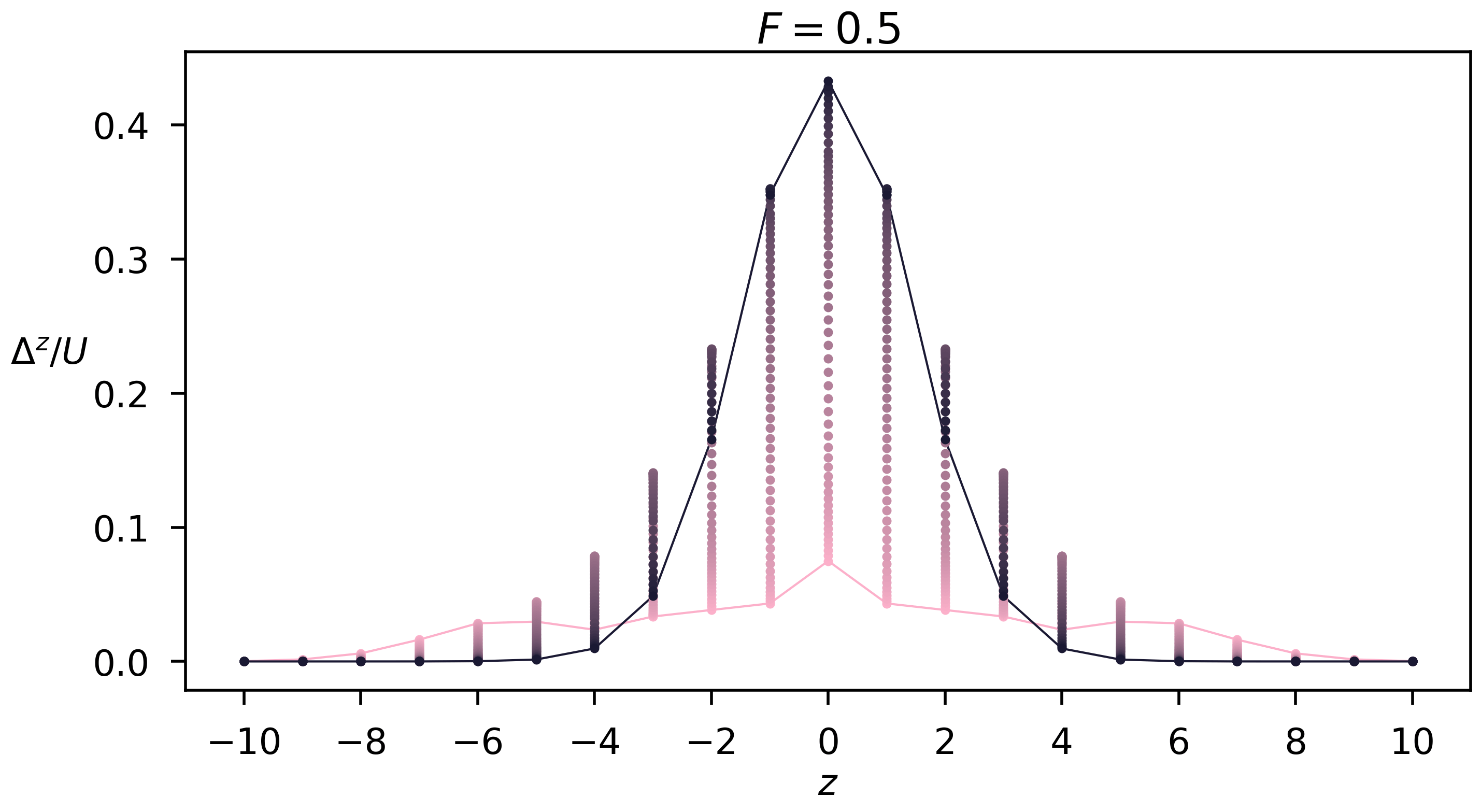}\\
    \includegraphics[width=8.6cm]{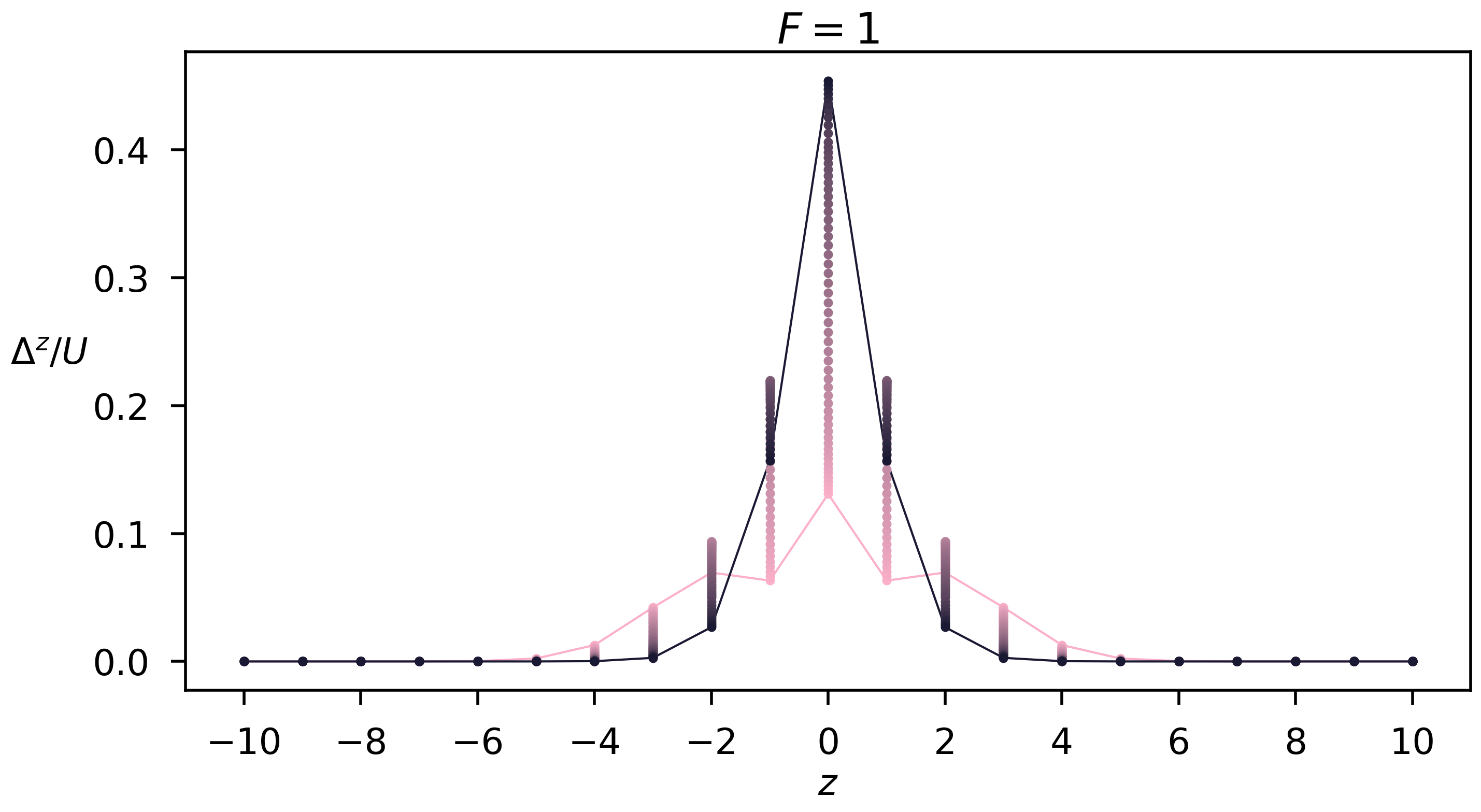}\\
    \raggedleft\includegraphics[width=7.6cm]{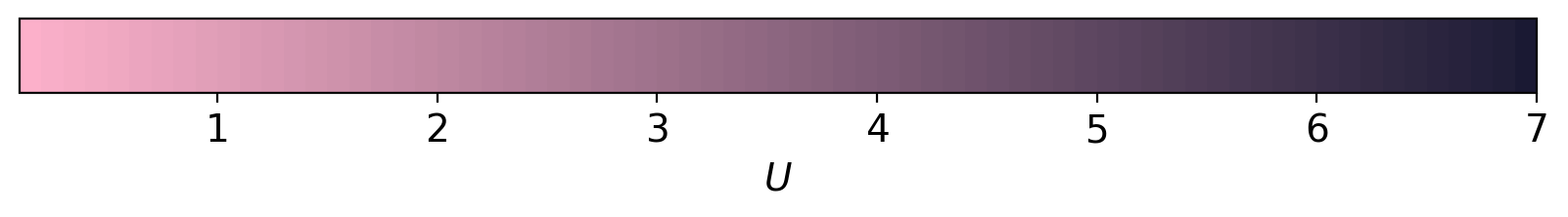}
    \caption{
        (Color online) Values of \(\Delta^z/U\) for \(0.1\leq U\leq 7\) with \textbf{Top}: \(F = 0.5\) and \textbf{Bottom}: \(F=1\), where light (dark) colors represent weaker (stronger) interaction strength.
        Lines are shown for \(U=0.1\) and \(U=7\).
        \(\Delta^z(z=0)/U\) increases with stronger interaction.
        Additionally, the spread becomes narrower as \(F\) and the band gap increase, corresponding directly to the partially filled bands in Fig.~\ref{fig:dens_square}.
    }
    \label{fig:delta_square}
\end{figure}

The first ingredient for superconductivity is the existence of the Cooper pairs, quantified by the pairing order parameter, \(\Delta^{z}/U\), and the single particle correlation length, \(\xi\), characterizing the spatial extent of the pair.
For dispersive bands in 2D, the critical temperature of pairing is of the order of \(k_B T_c \sim \Delta(T=0)\sim e^{-a/U}\), but it has been shown that \(k_B T_c \sim \Delta(T=0)\sim U\) for flat  bands~\cite{peotta2015superfluidity,heikkila2016flat,kopnin2011high,peri2021fragile,heikkila2011flat,verma2021optical}.
We show in Figs.~\ref{fig:dens_square} and~\ref{fig:delta_square} the \(z\)-band filling, \(\rho^z\), and pairing order parameter, \(\Delta^z/U\), at \(F=0.5\) and \(F=1\), with a range of attraction strengths \(0.1\leq U\leq7\).
While the filling on the \(z=0\) band is \(\rho^{z=0}=0.5\) for all values of \(U\) and \(F\) (as per our choice of the chemical potential), the spread of partially filled bands (\(0<\rho^z<1\)) narrows with increasing DC field and attraction strength (Fig.~\ref{fig:dens_square}).
Since \(\Delta^z/U\) is nonzero only for bands with partial filling, the number of bands with finite and large \(\Delta^z/U\) also decrease, evident in the narrowing peak about \(z=0\) in Fig.~\ref{fig:delta_square}.
Due to the larger band gap and the faster decay of the wavefunction as \(F\) increases, the number of partially filled \(z\)-bands decreases, even at finite interaction.
On the other hand, as \(U\) increases, the Cooper pairs are more tightly bound, increasing the \(\Delta^{z=0}\) peak, but narrowing the spread of the wavefunction in the \(\mathbf{e_z}\) direction.

\begin{figure}
    \centering
    \includegraphics[width=7.6cm]{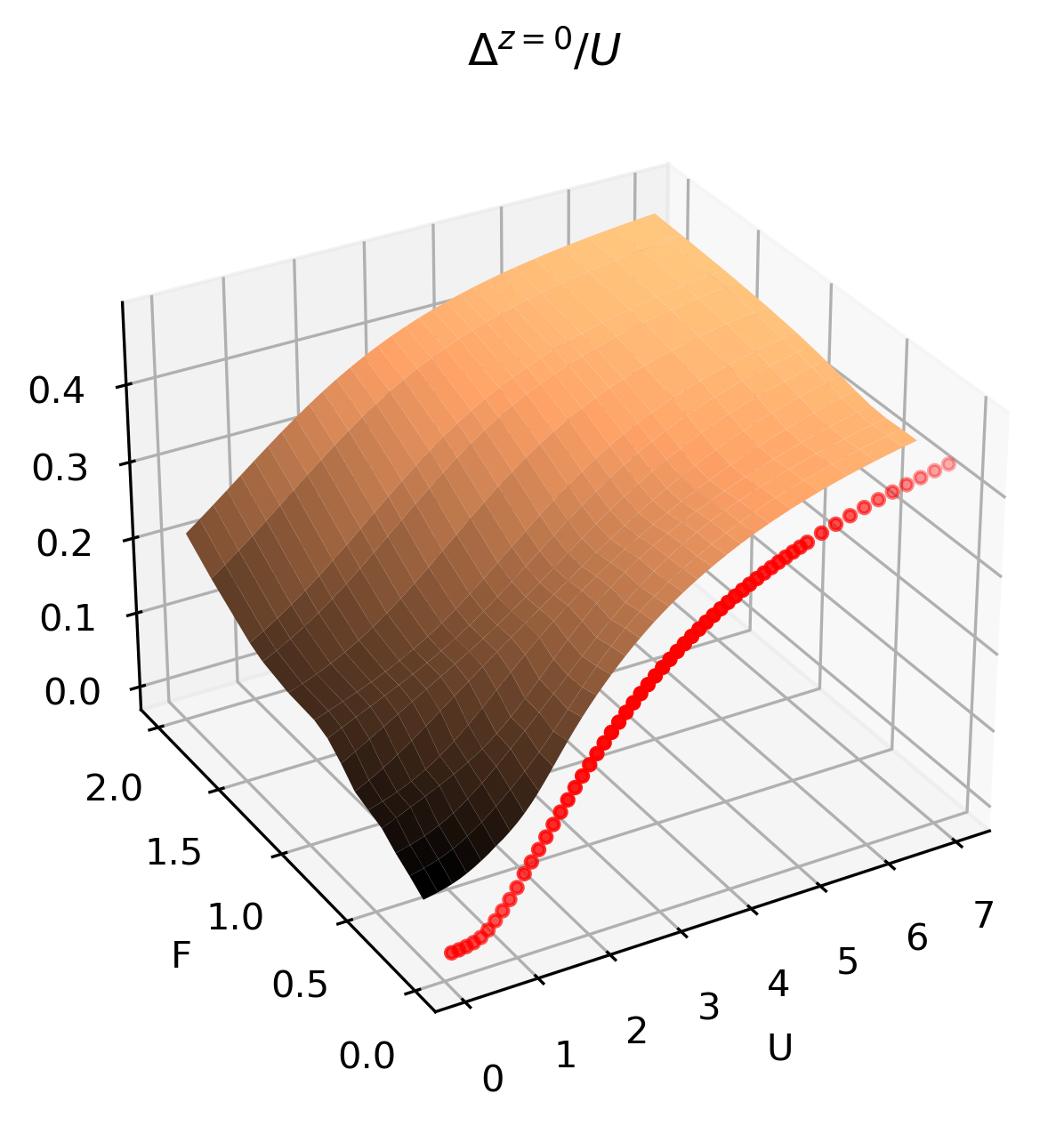}
    \caption{
        (Color online) The values of \(\Delta^{z=0}/U\) on the half-filled \(z=0\) band as a function of \(F\) and \(U\).
        The red points are \(\Delta/U\) of the half-filled square lattice with a dispersive band, when \(F=0\).
    }
    \label{fig:delta_z0}
\end{figure}

For these values of the DC field, \(F=0.5\) and \(F=1\), the pairing order parameter, \(\Delta^{z=0}/U\), on the half-filled band is finite for any weak interaction and increases with \(|U|\), which is usually observed for interacting flat band systems~\cite{chan2022pairing}.
In contrast to dispersive bands, in flat band systems \emph{with CLS}, the Cooper pair sizes are smaller than a single lattice spacing~\cite{chan2022pairing,chan2022designer} for \emph{any} value of \(U\),
thus there is no BCS-BEC crossover, and the pairing order parameter remains large even for weak interaction.
However, here, as both the DC field and \(U\) are further decreased, the pairing order parameter is substantially suppressed (Fig.~\ref{fig:delta_z0}).
In the limit \(F\rightarrow 0\), the pairing order parameter of the half-filled square lattice with a dispersive band is recovered on the \(z=0\) flat band, as the single particle wavefunction is increasingly spread out across the square lattice, even in the presence of a DC field.

\begin{figure}
    \includegraphics[width=8.6cm]{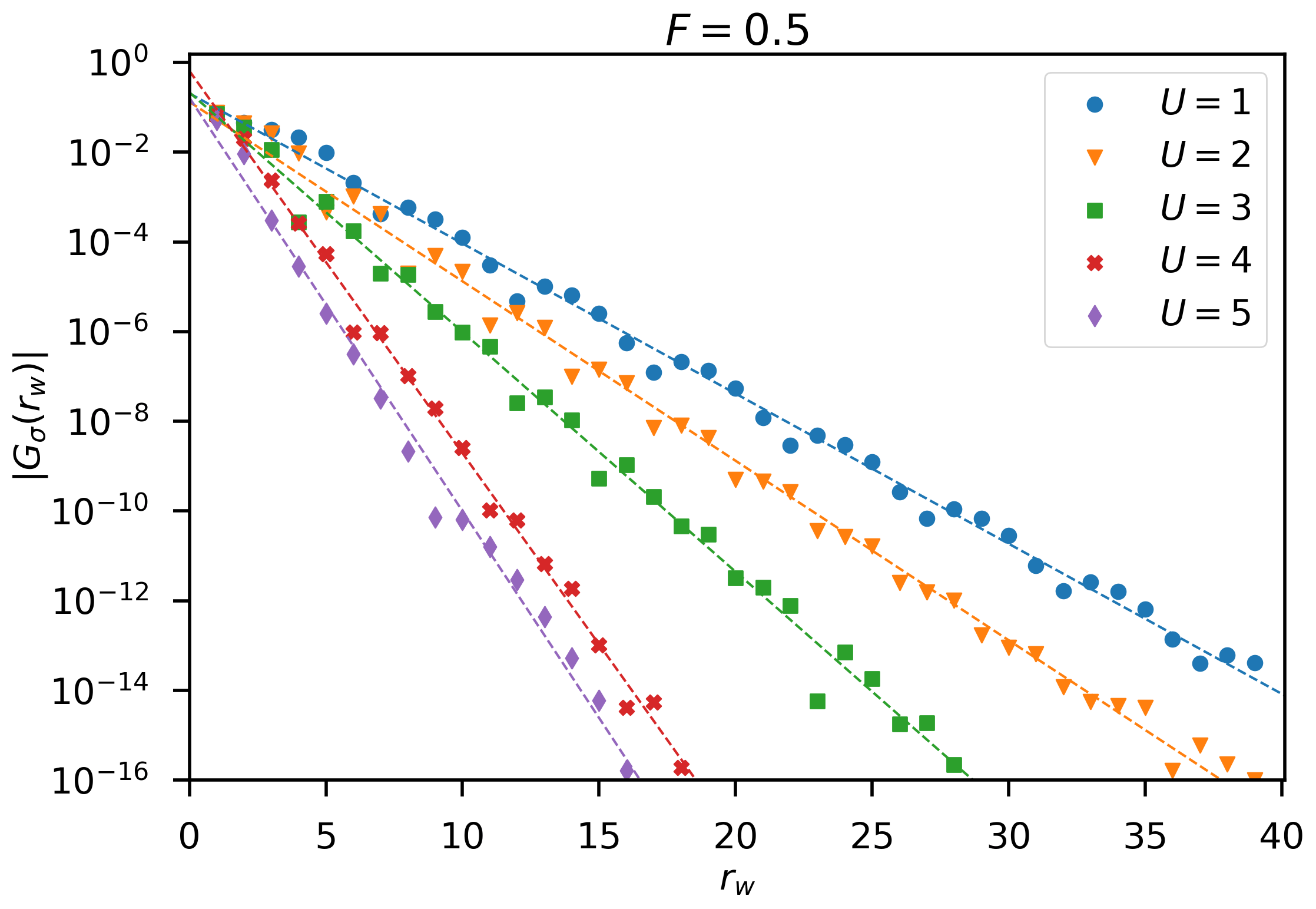} \\
    \includegraphics[width=8.6cm]{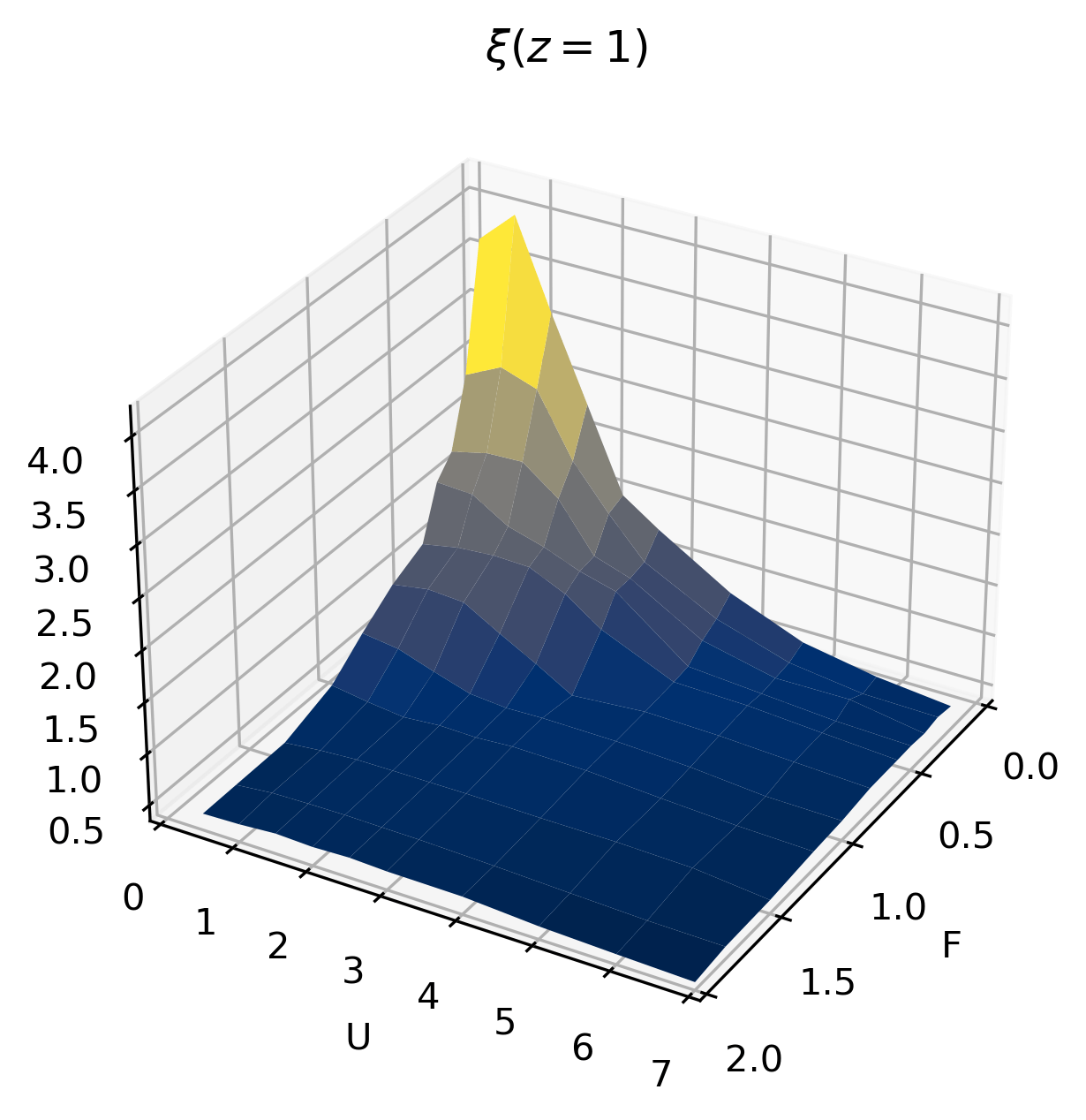}
    \caption{
        (Color online) \textbf{Top}: Correlation function on the 2D Wannier-Stark lattice decays exponentially along the \(\mathbf{e_w}\) direction, shown for \(F=0.5\).
        \textbf{Bottom}: Correlation length as a function of \(U\) and electric field strength in the \(\mathbf{e_w}\) direction.
        For sufficiently strong DC fields (\(F>1\)), the correlation length saturates to a value less than one lattice spacing as \(U\rightarrow 0\).
    }
    \label{fig:wcorr}
\end{figure}

\begin{figure}
    \includegraphics[width=8.6cm]{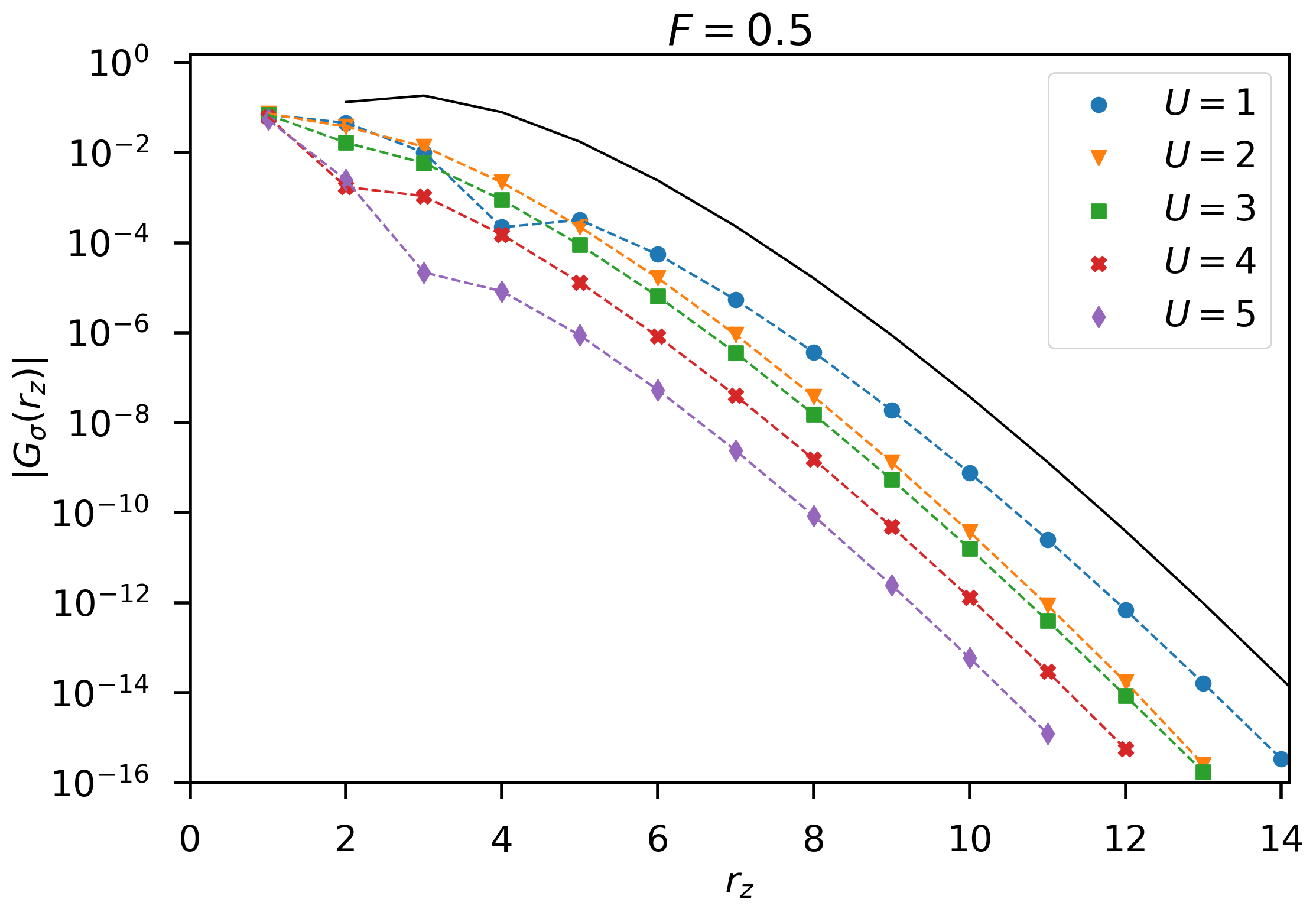} \\
    \includegraphics[width=8.6cm]{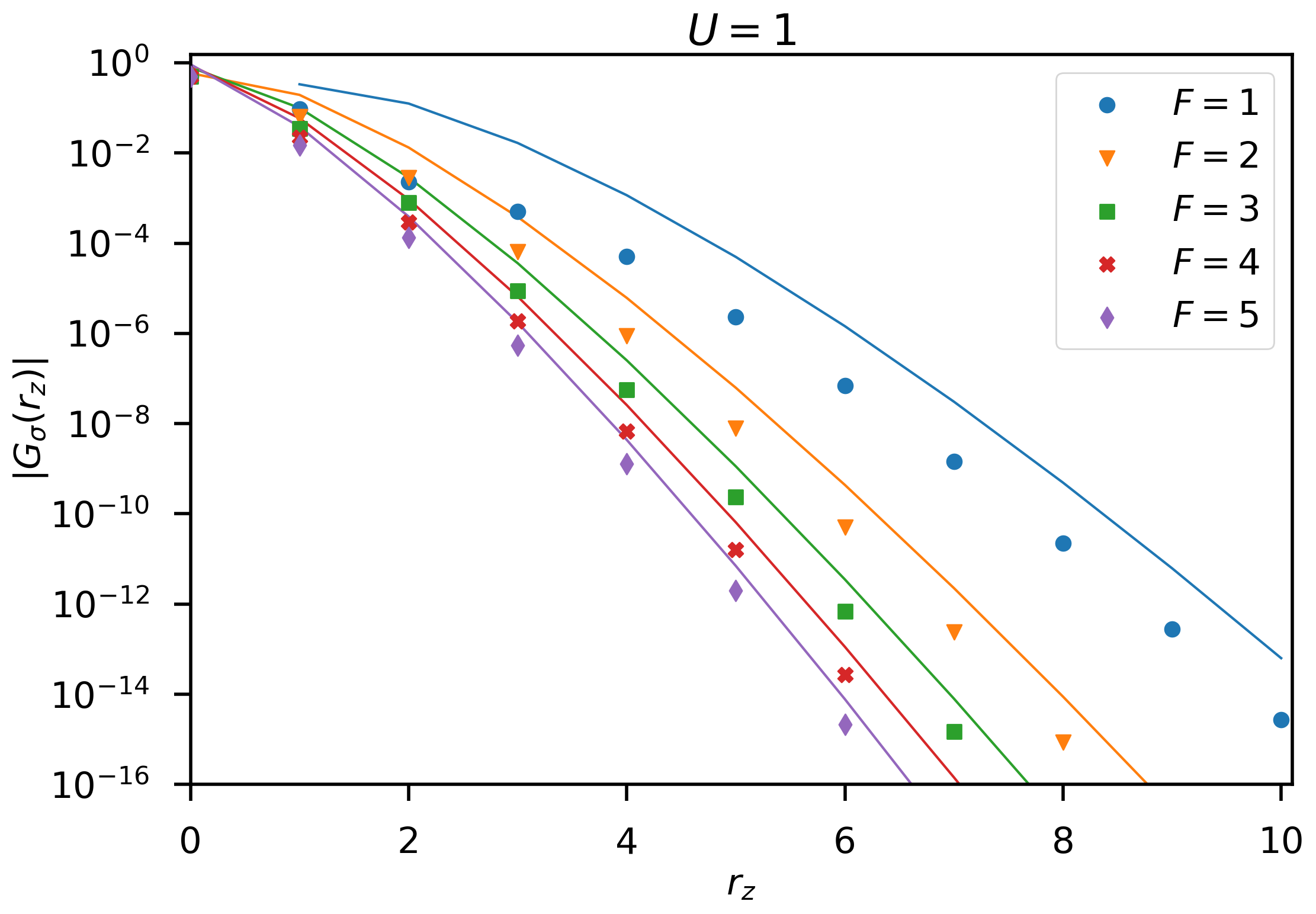}
    \caption{
        (Color online) The correlation function on the 2D Wannier-Stark lattice in the \(\mathbf{e_z}\) direction, \(\abs{G_\sigma(r_z)}\).
        \textbf{Top}: At \(F=0.5\) and several \(1\leq U\leq 5\).
        The black solid line is the non-interacting wavefunction at \(F=0.5\), \(\Psi(r_z,w=0)=J_{-r_z}(-4)J_{r_z}(4)\) (see Eq.~\eqref{eq:wavefunctions}).
        \textbf{Bottom}: At \(U=1\) and several \(1\leq F\leq 5\).
        The solid lines correspond to the non-interacting wavefunctions as defined in Eq.~\eqref{eq:wavefunctions}.
    }
    \label{fig:zcorr}
\end{figure}

This is an interesting feature of Wannier-Stark flat bands that is distinct from flat bands with CLS and can be understood as follows.
As \(F\) is decreased, the wavefunction becomes less localized, while the existence of flat bands persists. 
Figure~\ref{fig:wcorr} (top) shows the single particle correlation function as a function of distance along the flat band at \(z=1\)~\footnote{The correlation function along the diagonal on the \(z=0\) band is \(0\) because of the global and local (within the flat band) half-filling.  
This is due to the Fermi surface of the square lattice at half-filling and does not mean that the particles are uncorrelated.  }, \emph{i.e.} transverse to the field, and (bottom) the correlation length for various values of \(U\) and \(F\).
It is clear in the top panel that this correlation function decays exponentially, \(G_\sigma(r_w)\sim e^{-r_w\sqrt{2}/\xi}\), thus yielding the correlation length, \(\xi\), which characterizes the size of the pair.
The \(\sqrt{2}\) factor in the exponent arises from the distance between consecutive sites in the \(\mathbf{e_w}\) direction.  
At fixed \(F\), the decay is faster for larger \(U\) resulting in smaller pairs with smaller values of \(\xi\).
The bottom panel of Fig.~\ref{fig:wcorr} shows the correlation length in the \(\mathbf{e_w}\) direction as a function of \(F\) and \(U\). 
It is clear, then, that at small values of \(U\) and \(F\), the correlation length exceeds the lattice spacing, and that,
therefore, the pair extends over several unit cells in the transverse direction.
In addition, as \(F\) and/or \(U\) decrease, the pair wavefunction also spreads out along the direction of the DC field, \(\mathbf{e_z}\) as seen in Fig.~\ref{fig:zcorr}.
However, the decay of the wavefunction in the direction of the DC field is faster than exponential and is essentially given by a rescaling of the \(U=0\) wavefunction, Eq.~\eqref{eq:wavefunctions}.
This spread of the wavefunction along \(\mathbf{e_z}\) and \(\mathbf{e_w}\) as \(F\) and \(U\) decrease causes the order parameter, \(\Delta^z/U\), to decrease as seen in Fig.~\ref{fig:delta_z0}.

\begin{figure}
    \includegraphics[width=4.3cm]{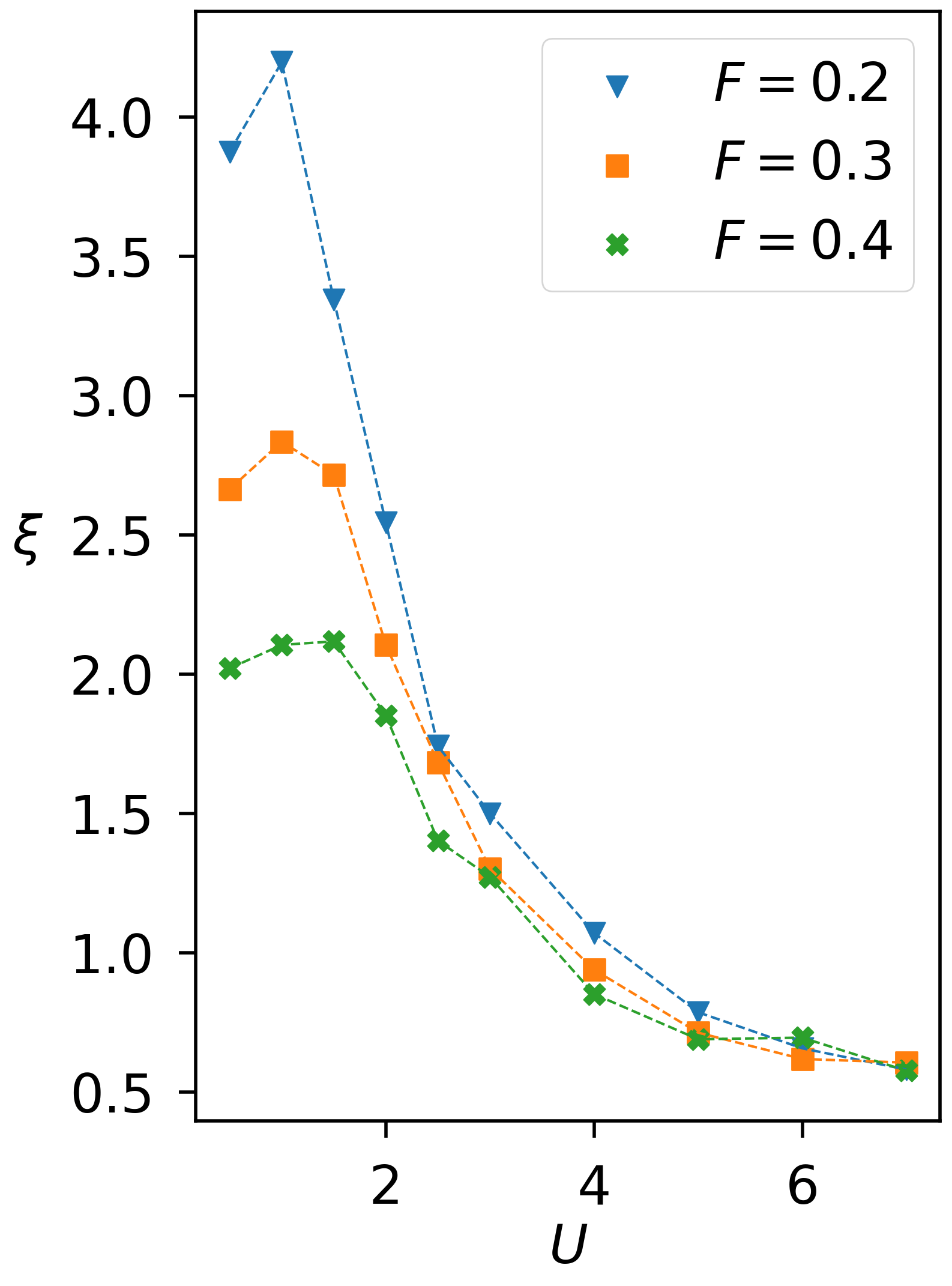}\includegraphics[width=4.3cm]{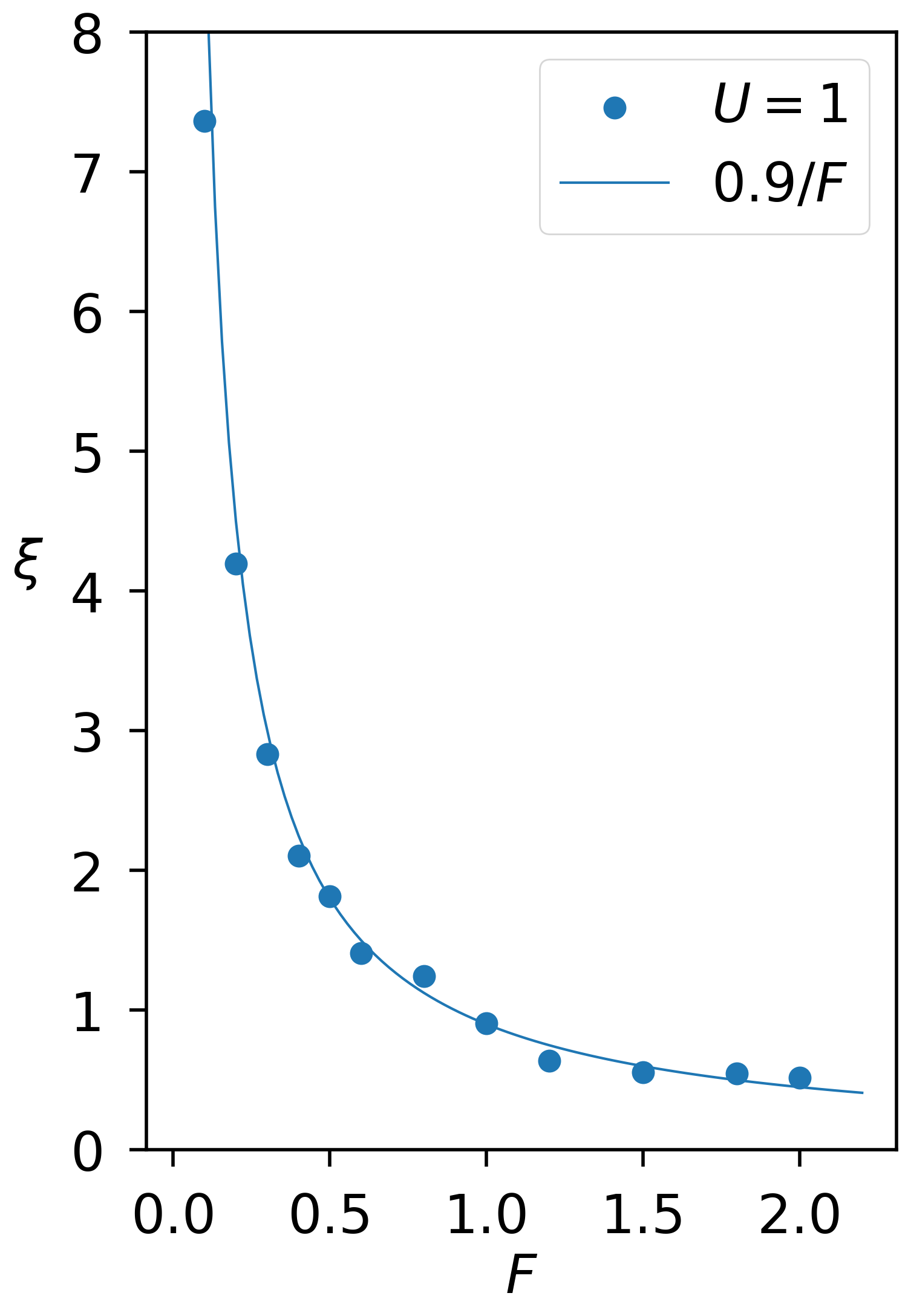}
    \caption{
        (Color online) Correlation length along the \(\mathbf{e_w}\) direction on the 2D Wannier-Stark lattice.
        \textbf{Left}: fixed \(F\) as a function of \(U\) for three values of \(F\).
        The lines are included to guide the eye. 
        \textbf{Right}: fixed \(U\) as a function of \(F\).
    }
    \label{fig:wcorrlengthcuts}
\end{figure}

It is interesting to elaborate some more on the behavior of the correlation functions along and transverse to the applied DC field as \(F\) or \(U\) decrease.
First, we emphasize that for any nonzero value of \(F\), the flat bands exist.
At \(U=0\), the wavefunction is given by Eq.~\eqref{eq:wavefunctions} and is the same along and transverse to the applied DC field: it spreads isotropically as \(F\) decreases, and it decays faster than exponentially.
When \(U > 0\), the wavefunction becomes a scaled Bessel function in the direction of the applied field, as seen in Fig.~\ref{fig:zcorr}, and therefore still decays faster than exponential.
However, in the transverse direction it decays exponentially (Fig.~\ref{fig:wcorr}).
Fixing \(U\) at a small value and calculating \(\xi\) as a function of \(F\), we find that \(\xi\) increases rapidly as \(F\) decreases, Fig.~\ref{fig:wcorr} (bottom).
We show in Fig.~\ref{fig:wcorrlengthcuts} (right) a cut along \(U=1\) demonstrating this rapid increase which appears to be a power law divergence \(\xi \sim 1/F\).
Going to an even smaller values of \(F\) was too challenging, since it required extremely large system sizes.
On the other hand, fixing \(F\) at some small value and taking \(U\rightarrow 0\), the behavior is different: 
at first \(\xi\) increases, but then eventually it decreases, Fig.~\ref{fig:wcorrlengthcuts} (left).
This is because as \(U\) vanishes, the wavefunction reverts to its non-interacting limit and decays faster than exponentially in all directions.

The second essential ingredient for superconductivity is for the pairs to exhibit phase stiffness which is manifested by nonzero superfluid weight, Eq.~\eqref{eq:Dscompare}.
The spread and density overlap of the single particle flat band wavefunctions were previously found to have direct implications on the superfluid weight when \(U\neq 0\)~\cite{chan2022designer,tovmasyan2016effective}.
We therefore investigate the dependence of \(D_s\) on \(F\) and \(U\).

\begin{figure}
    \includegraphics[width=7.6cm]{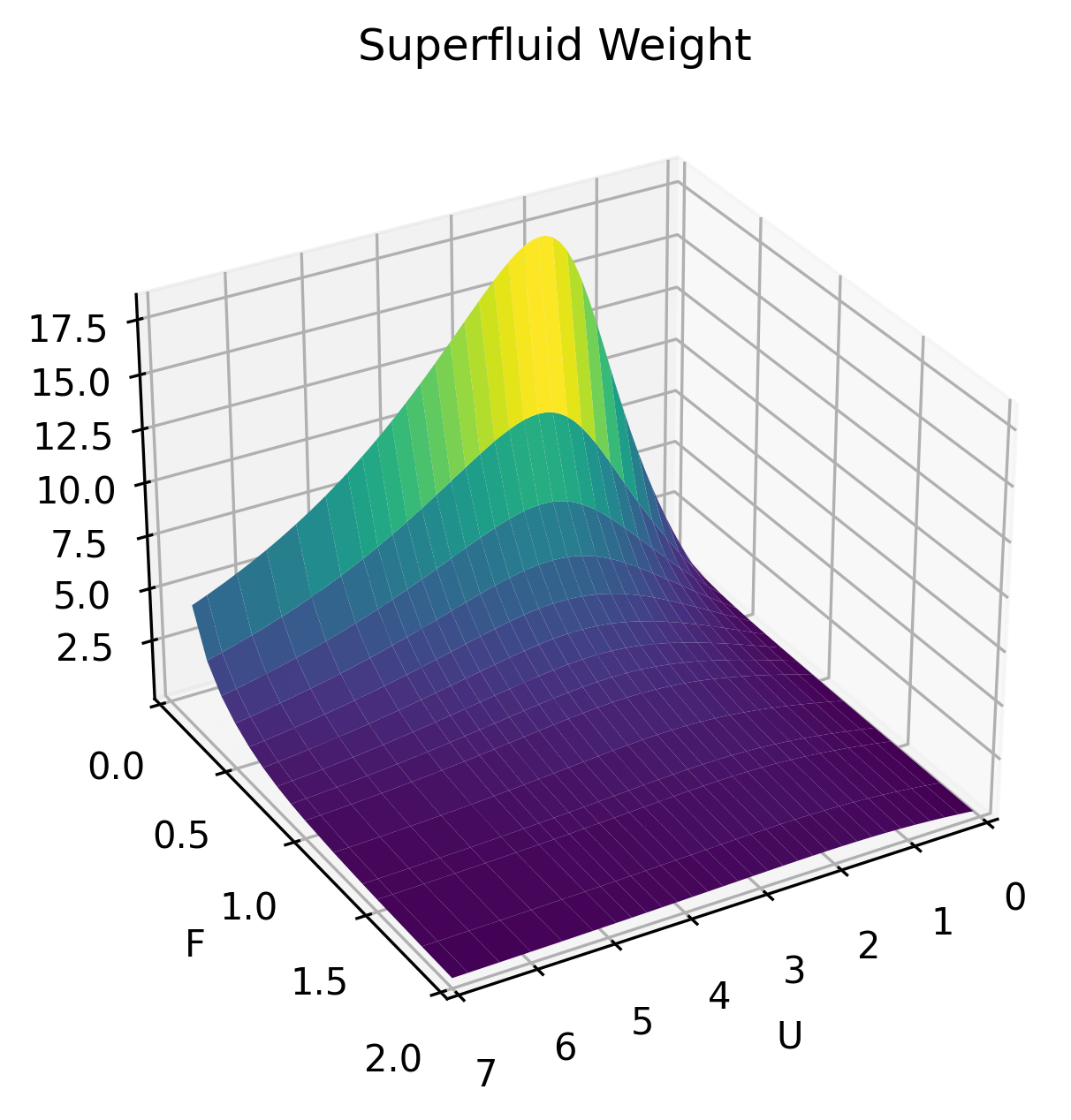}
    \caption{
        (Color online) Superfluid weight on the 2D Wannier-Stark lattice along the \(\mathbf{e_w}\) direction as a function of DC field and Hubbard interaction strengths.
        (Details in Appendix~\ref{appendix:Ds}.)
    }
    \label{fig:Ds}
\end{figure}

We observe in Fig.~\ref{fig:Ds} (and Fig.~\ref{fig:Ds2Dcuts} in Appendix~\ref{appendix:Ds}) that for each value of \(F\), the superfluid weight at first increases linearly with \(U\), 
reaches a peak at around \(U\approx 2\), and then decreases. The linear dependence on \(U\), for small \(U\), has been previously shown to be caused by the flat band, with nonzero minimal quantum metric~\cite{chan2022designer,tovmasyan2018preformed,tovmasyan2016effective,huhtinen2022revisiting,thumin2023flat}.
Note that in these cases the CLS have nonzero density overlap.
For quasi one-dimensional non-topological bands, \emph{i.e.} with vanishing minimal quantum metric, the CLS are localized within one unit cell resulting in a suppression of superfluid transport leading to \(D_s\propto U^\alpha\) with \(\alpha > 2\).
Note that in these cases the CLS have zero density overlap.
Interestingly, the quasi one-dimensional flat bands in our case, resulting from the applied DC field, are non-topological yet yield \(D_s \propto U\) for small \(U\).
The reason for this might be that in the current system, the localized state is not compact, the wavefunction, Eq.~\eqref{eq:wavefunctions}, extends over more than one unit cell and results in nonzero density overlap, which then may lead to a linear dependence on \(U\).
At very large \(U\), the physics is very well described by an effective extended hard-core Bose-Hubbard model which exhibits a decreasing \(D_s\) as \(U\) increases~\cite{mondaini2018pairing,chan2022designer,chan2022pairing,emery1976theory,thumin2023flat}.
These two branches of \(D_s\), linear increase at low \(U\) and decrease at large \(U\), meet in the intermediate \(U\) regime where the physics is dominated by band mixing and \(D_s\) displays a maximum.

We expect this maximum value to be of the order of the relevant band width of the unbiased noninteracting lattice which is proportional to the hopping \(t=1\).
For a fixed nonzero DC field value, the single particle wave function extension along the \(z\)-direction is proportional to \(\frac{t}{F}\).
Let us assume that \(F < t\) which corresponds to the most interesting regime of large superfluid weight.
For \(U < F\) we expect very weak superfluid features, as discussed above.
Increasing \(U\) will couple more and more flat bands and thus increase the superfluid weight.
When \(U\) reaches a value around \(U\sim t\) all bands within the reach of one eigenfunction extent along the \(z\)-direction are coupled.
Further increase of \(U\) will not lead to a further increase of the weight, but instead lead smoothly into the regime of hard core bosons which means instead a decrease of the superfluid weight.
Therefore, \(U_\mathrm{max} \sim t\), and we can expect that the value of the superfluid weight at the maximum will increase as \(F\) is further decreased since more and more flat bands participate.
For the case \(F > t\), we expect \(U_\mathrm{max} \rightarrow 0\) and a suppression of superfluidity with increasing \(F\), since the density overlap between localized flat band states will quickly decrease.

Remarkably, as the strength of the DC field is decreased, the superfluid weight is enhanced dramatically, unlike the pairing order parameter, \(\Delta^z/U\), which decreases as discussed above.
This enhancement of \(D_s\) can be understood as follows: The wavefunction asymptotics of this Wannier-Stark system in the non-interacting limit is~\cite{mallick2021wannier}
\begin{equation}
    \abs{\Psi(r_z, w=0)}\approx\frac{1}{\abs{r_z}!^2}\abs{\frac{t}{F}}^{2\abs{r_z}}.
\end{equation}
Since any finite DC field breaks the periodicity in the \(\mathbf{e_z}\) direction, a weak DC field produces multiple flat bands in the \(\mathbf{e_w}\) direction with diminished gaps of \(E_\mathrm{gap}=F\) between consecutive \(z\)-bands.  
Additionally, the wavefunction in the \(\mathbf{e_z}\) direction broadens \(\sim \frac{t}{F}\), directly increasing the number of partially filled \(z\)-bands which can participate in supercurrent transport.
Consequently, the smaller, but nonzero, \(F\) is, the more closely packed the flat bands become, the more of them are covered by the spreading wavefunction and are doped (see Fig.~\ref{fig:dens_square}).
This results in the observed improvement of \(D_s\).
We emphasize that while the pairing order parameter is a local quantity and converges to the dispersive square lattice limit as \(F \rightarrow 0\) and the wavefunction spreads, the superfluid weight describes the dissipationless transport of the entire system and depends on the number of bands contributing to flat band superconductivity.
Moreover, the system is effectively quasi one-dimensional in the presence of a DC field, thus the square lattice limit of \(D_s\) will \emph{not} be recovered even as \(F\rightarrow 0\), and is in fact dramatically enhanced with a weak DC field strength.

\begin{figure}
    \includegraphics[width=8.6cm]{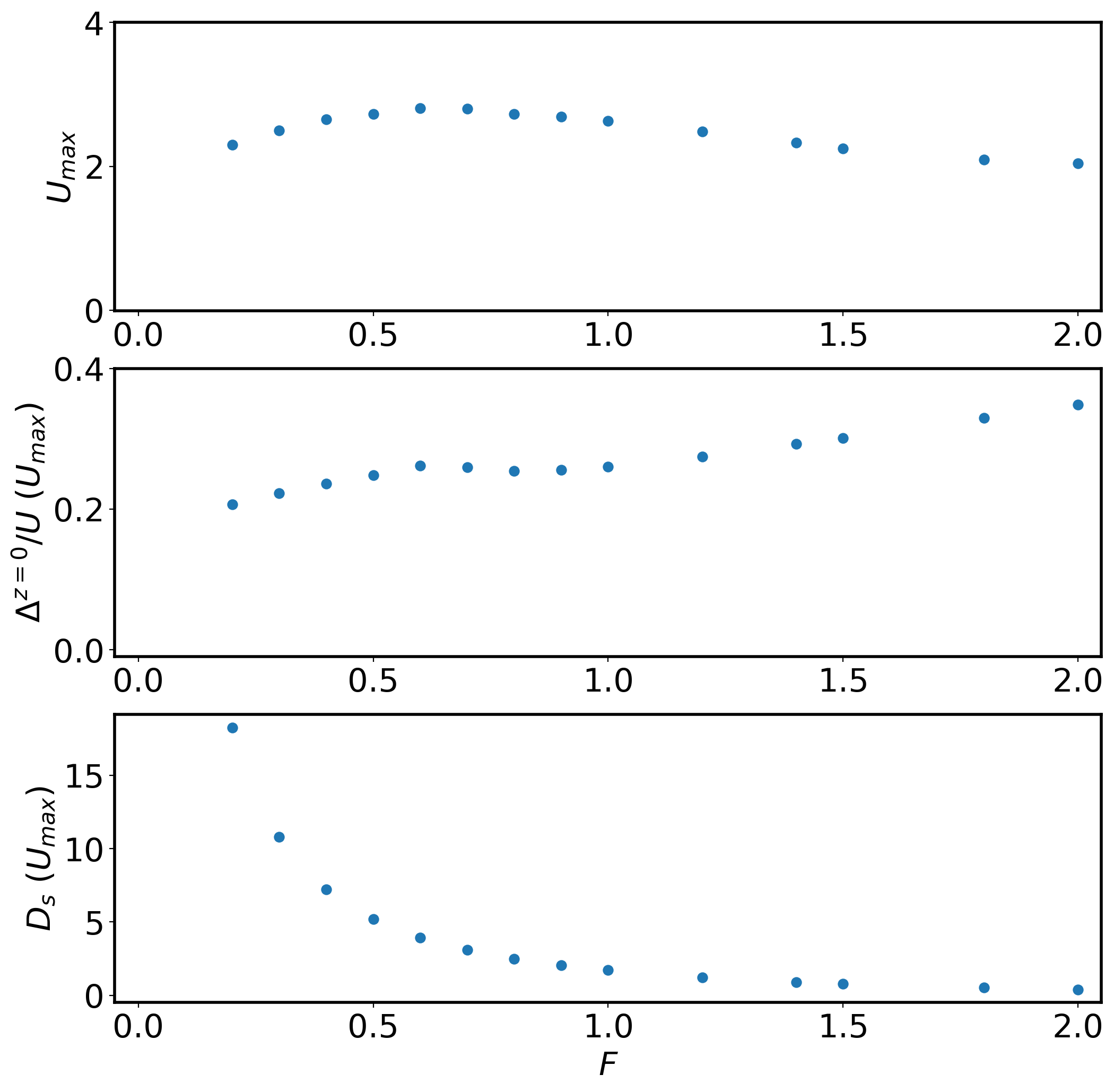}
    \caption{
        (Color online) 
        \textbf{Top}: The value of \(U_\mathrm{max}\) corresponding to the peak in \(D_s\) as a function of \(F\).
        \textbf{Middle}: Pairing order parameter at \(U_\mathrm{max}\) as a function of \(F\).
        \textbf{Bottom}: The maximum attainable superfluid weight as a function of \(F\).
    }
    \label{fig:Umax}
\end{figure}

It is interesting to study the dependence of the maximum attainable superfluid weight, \(D_s\), as a function of the DC field strength.
We observe in Fig.~\ref{fig:Umax}, that the interaction strength and pairing order parameter which correspond to the maximum \(D_s\) do not vary significantly as the DC field strength is changed.
As previously explained, \(U_\mathrm{max}\) is attained when all partially-filled bands are coupled, at \(U \sim t\).
However, due to the increasing spread of the wavefunction and narrowing of band gaps, the maximum \(D_s\) grows and diverges as \(F\rightarrow 0\). 

For superconductivity to remain robust at finite temperature, both \(D_s\) and \(\Delta^z/U(T=0)\) need to be nonzero and large.
This demands a balance between increasing both the number of contributing flat bands (small \(F\)) and the pairing order parameter (intermediate \(F\)) at weak interaction.
We, therefore, believe that an intermediate \(F\) is optimum for significant superfluid weight while maintaining pair coherence and enhancing the critical temperature of pairing.

\subsection{3D Cubic Lattice}

For completeness, we extend the study to the 3D Wannier-Stark lattice defined in Eq.~\eqref{eq:3DHam}.
By applying a DC field in the \(\mathbf{e_\gamma}\) direction, we have \(F=\frac{aE}{\sqrt{3}}\) with \(E\) the DC field strength.
The flat bands which result are perpendicular to \(\mathbf{e_\gamma}\) and have an energy gap of \(F\) between consecutive planes (Fig.~\ref{fig:WannierStarkBands}).
Here, the \(\gamma=0\) band is fixed at half-filling.

\begin{figure}
    \includegraphics[width=7.6cm]{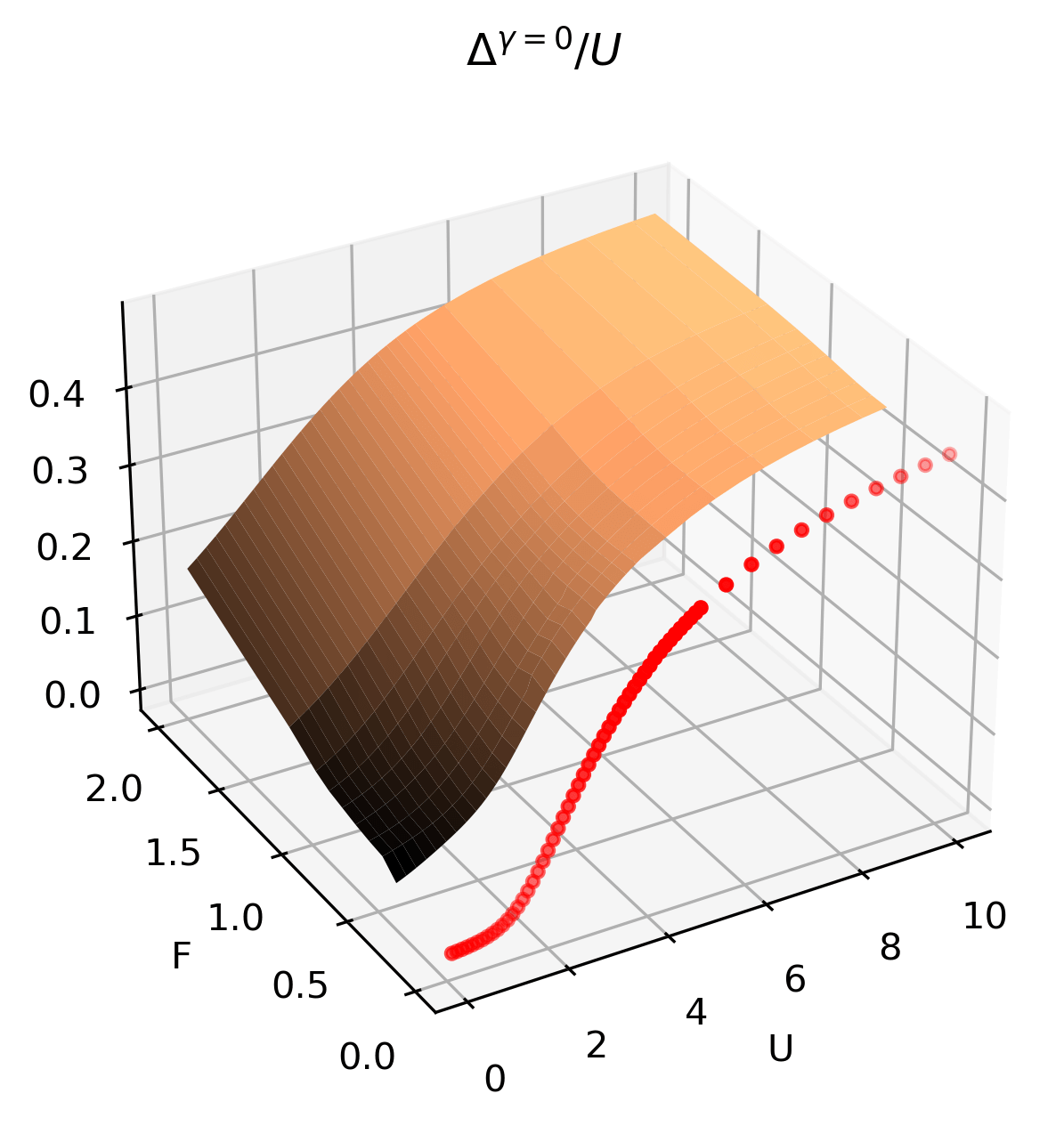}
    \caption{
        (Color online) The pairing order parameter \(\Delta^{\gamma=0}/U\) as a function of the DC field and interaction strengths on the half-filled \(\gamma=0\) band for the cubic lattice.
        The red points are the pairing order parameter on a half-filled cubic lattice with a dispersive band, when \(F=0\).
    }
    \label{fig:Deltas3Dcubic}
\end{figure}

Overall the results are similar to the 2D case.
As before, we observe a decrease of the pairing order parameter, \(\Delta^{\gamma=0}/U\), as both the DC field \(F\) and attraction strengths \(U\) decrease, due to the delocalization of the single particle wavefunction as \(F\) is tuned to zero.
However, at weak interaction, we still have a large and finite value for sufficiently strong \(F\)~(Fig.~\ref{fig:Deltas3Dcubic}).
In the limit \(F\rightarrow 0\), we recover the pairing order parameter of the cubic lattice shown in red points in Fig.~\ref{fig:Deltas3Dcubic}.

\begin{figure}
    \includegraphics[width=7.6cm]{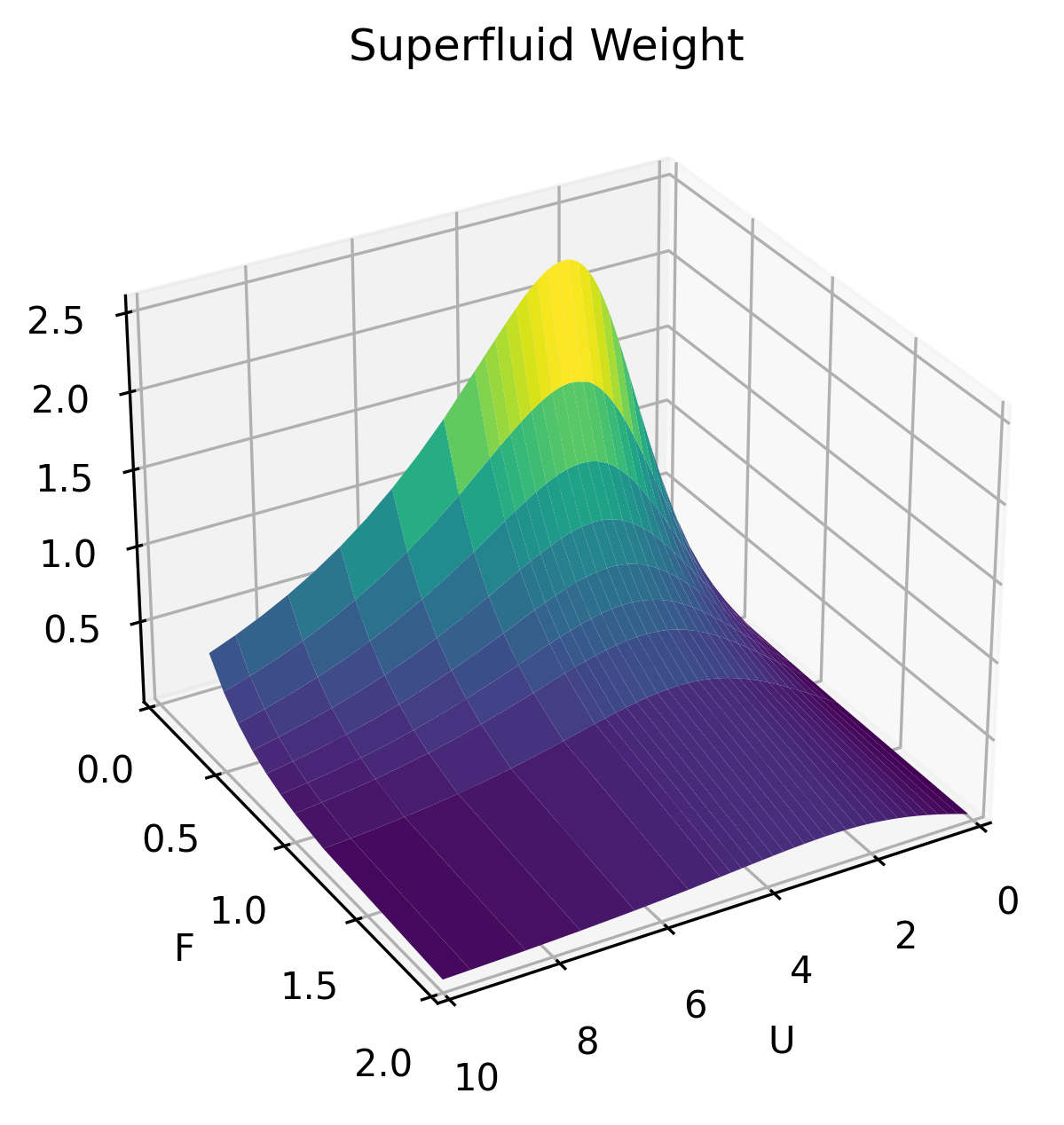}
    \caption{
        (Color online) The superfluid weight perpendicular to \(\mathbf{e_\gamma}\) in the 3D Wannier-Stark lattice as a function of \(U\) and \(F\).
        (Details in Appendix~\ref{appendix:Ds}.)
    }
    \label{fig:Ds3D}
\end{figure}

Figure~\ref{fig:Ds3D} illustrates the dependence of superfluid weight, \(D_s\), on the DC field and interaction strength for the 3D Wannier-Stark lattice.
Again, there is a strong enhancement of \(D_s\) when the DC field strength is weak, which produces multiple partially-filled flat bands with small band gaps.

\section{Conclusions}
\label{sec:conclusions}

In this study, we characterize in detail the superconducting properties of interacting fermions in Wannier-Stark flat band lattices obtained by applying a DC field along a diagonal of the square lattice.
Superconducting transport has two main ingredients, the formation of Cooper pairs which become a condensate (or quasi-condensate) and their nonzero superfluid weight due to the phase stiffness of pairs.
We calculated the pairing order parameter, correlation functions, and the superfluid weight while systematically varying the DC field and interaction strength, and uncovered the dependence of the superconducting properties on these parameters.

We showed that, unlike flat band systems with a compact localized state, CLS, here transport is supported cooperatively by multiple flat bands as the spatial extent of the wavefunction spreads across several \(z\) coordinates corresponding directly to flat \(z\)-bands (or \(\gamma\)-bands in 3D).
Therefore, the physics is different from flat band systems with CLS and nonzero minimal quantum metric where superconductivity at weak interaction is governed by the eigenstates and quantum geometry of the flat band.
Despite the trivial topology of this system, we have demonstrated that it supports flat  band superconductivity due to the non-compact nature of the localized state.
Furthermore, the superconductivity can be enhanced easily on these Wannier-Stark flat bands by tuning the amplitude of the applied DC field.
Reducing the DC field strength increases the number of partially-filled flat band channels that contribute towards superconductivity and the overlap of the single particle wavefunctions.
At weak coupling, the pairing order parameter is small as \(F\) decreases, thus an intermediate DC field strength is desirable to optimize both the superfluid weight and the critical temperature.

The results presented would be qualitatively similar for any commensurate direction that the DC field is oriented in, and on all Wannier-Stark flat band systems, independent of the underlying Bravais lattice.
We propose that the experimental realization of Wannier-Stark flat band systems is advantageous as they require less fine tuning but can similarly result in flat band superconductivity.
For example, it is possible to study this system in experiments with ultra-cold atoms loaded in optical lattices.
In this situation, the DC field is emulated by the gravitational field when the lattice is tilted at an  angle~\cite{anderson1998macroscopic,ivanov2008coherent,roati2004atom,sapienza2003optical}.
Moreover, there is no CLS, and the density overlap of the non-interacting wavefunctions can be controlled with the DC field strength, which increases the coherent transport of the Cooper pairs.

It would be insightful to analyze the many-body interactions and superconducting properties on non-Bravais Wannier-Stark lattices and topological flat Wannier-Stark bands, like the ones proposed in Refs.~\onlinecite{mallick2022antipt}
and~\onlinecite{kolovsky2018topological}.
Since the effective dimension of the superconducting channel is \(D-1\), it would also be interesting to explore the effect of disorder on Wannier-Stark flat bands.

\begin{acknowledgments}
    S.M.C. is supported by a National University of Singapore President's Graduate Fellowship and the Center for Theoretical Physics of Complex Systems, Institute for Basic Science (PCS-IBS) Visitors Program.
    S.M.C. would like to thank PCS-IBS in the Republic of Korea for their kind hospitality and stimulating research environment.
    AA and SF acknowledge financial support from the Institute for Basic Science (IBS) in the Republic of Korea through the project IBS-R024-D1.
\end{acknowledgments}

\appendix

\section{Mean Field Approximation on Wannier-Stark Flat Bands}
\label{appendix:multibandMF} 

To derive the mean-field expression of the interaction term in Eq.\eqref{eq:Hint_2D} of the main text, we first recognize that sites on equivalent \(z\)-bands are identical, but the mean field parameters should depend on the index \(z\).
In other words, the pairing and filling on different \(z\)-bands should be distinct.
We can then write a trial Hamiltonian for the interaction part,
\begin{equation}
\begin{aligned}
    H_{I,\mathrm{trial}}=& - \frac{1}{2}\displaystyle\sum_{z,w}\left(\Delta^z
    c_{z,w,\downarrow}^\dagger c_{z,w,\uparrow}^\dagger + \Delta^{z*}
    c_{z,w,\uparrow}^{\phantom \dagger}c_{z,w,\downarrow}^{\phantom \dagger}\right) \\
    &-\frac{U}{2}\displaystyle\sum_{z,w}\left(\rho^z_\uparrow c_{z,w,\downarrow}^\dagger c_{z,w,\downarrow}^{\phantom \dagger} + \rho^z_\downarrow c_{z,w,\uparrow}^\dagger c_{z,w,\uparrow}^{\phantom \dagger} \right)
\end{aligned}
\end{equation}
with the mean-field parameters \(\Delta^z\), \(\rho^z_\uparrow\), \(\rho^z_\downarrow\) to be defined.

We write the Gibbs-Bogoliubov inequality at \(T=0\)~\cite{kuzemsky2015variational},
\begin{equation}
    \begin{aligned}
        \langle H_I \rangle \leq & -
        \frac{U}{2}\displaystyle\sum_{z,w} \left\langle
        c_{z,w,\uparrow}^\dagger c_{z,w,\uparrow}^{\phantom \dagger}
        c_{z,w,\downarrow}^\dagger c_{z,w,\downarrow}^{\phantom
          \dagger} \right\rangle_\mathrm{trial} \\ 
        &+
        \frac{1}{2}\displaystyle\sum_{z,w}\left\langle\left(\Delta^z
        c_{z,w,\downarrow}^\dagger c_{z,w,\uparrow}^\dagger +
        \Delta^{z*} c_{z,w,\uparrow}^{\phantom \dagger}
        c_{z,w,\downarrow}^{\phantom \dagger}\right)
        \right\rangle_\mathrm{trial} \\ 
        &+\frac{U}{2}\displaystyle\sum_{z,w}\left\langle\left(
        \rho^z_\uparrow
        c_{z,w,\downarrow}^\dagger c_{z,w,\downarrow}^{\phantom \dagger} +
        \rho^z_\downarrow c_{z,w,\uparrow}^\dagger
        c_{z,w,\uparrow}^{\phantom \dagger} 
        \right)\right\rangle_\mathrm{trial}\\ 
        &+\left\langle H_{I,\mathrm{trial}}
        \right\rangle_\mathrm{trial} 
    \end{aligned}
\end{equation}
where \(\langle \dots \rangle_{trial}\) denotes expectation values with respect to the weight \({\rm e}^{-\beta H_{trial}}/Z_{trial}\).
Minimizing, we obtain the mean field interaction term and the mean-field parameters,
\begin{equation}
\begin{aligned}
     \langle \widetilde{H}_I \rangle =& \langle H_{I,\mathrm{trial}}
     \rangle +\frac{1}{2}L_w\displaystyle\sum_{z}
     \left(U\rho^z_{\uparrow}\rho^z_{\downarrow}
     +\abs{\Delta_{z}}^2/U\right) \\ 
     \rho^z_{\uparrow} =& \langle c_{z,w,\uparrow}^\dagger
     c_{z,w,\uparrow}^{\phantom\dagger} \rangle \\ 
     \rho^z_{\downarrow} =& \langle c_{z,w,\downarrow}^\dagger
     c_{z,w,\downarrow}^{\phantom\dagger}  \rangle \\ 
     \Delta^z_{\uparrow} =& \langle c_{z,w,\uparrow}
     c_{z,w,\downarrow} \rangle 
\end{aligned}
\end{equation}

With this approximation, we can write the Hamiltonian in the Bogoliubov-de Gennes (BdG) form in reciprocal space and diagonalize,
\begin{equation}
\begin{aligned}
    2(H-\mu N)=&\displaystyle\sum_{k} \Psi_{k}^\dagger \mathcal{M}(k)
    \Psi_{k}+ \mathrm{const.} \\
    =& \displaystyle\sum_{k}\Psi_{k}^\dagger
    \mathcal{P}(k)\mathbf{\Lambda}(k)\mathcal{P}^{-1}(k)\Psi_{k} +
    \mathrm{const.} \\ 
    =& \displaystyle\sum_{k}\Gamma_k^\dagger\mathbf{\Lambda}(k)
    \Gamma_k + \mathrm{const.}
\end{aligned}
\end{equation}
where \(\mathrm{const.} = L_w\displaystyle\sum_{z} \left(U\rho^z_{\uparrow}\rho^z_{\downarrow} +\abs{\Delta_{z}}^2/UFz-\mu-U\rho^z_\uparrow\right)\).
The Nambu spinor \(\Psi_k\) comprising the operators \(c_{z,w,\uparrow}^{\phantom\dagger} \) and \(c_{z,w,\downarrow}^\dagger\) can be identified to be 
\begin{equation}
    \Psi_k = \mathcal{P}(k)\Gamma_k
\end{equation}
where \(\Gamma_k\) consists of operators for quasiparticle excitation, \(\gamma_n\), \(n=1,2,\dots,2\lmax+1\), corresponding to positive eigenvalues and \(\gamma_n^\dagger\), \(n=2\lmax+2, 2\lmax+3, \dots,2(2\lmax+1)\) corresponding to negative eigenvalues.
The columns of \(\mathcal{P}(k)\) are eigenvectors, \(u_n\), of \(\mathcal{M}(k)\).
Recognizing that there are no quasiparticle excitations in the ground state, \(\langle \gamma_n^\dagger\gamma_n\rangle=0\), the only contributing terms are \(\langle \gamma_n \gamma_n^\dagger\rangle=1\) and we can determine the mean field parameters self-consistently with the eigenvectors of the BdG Hamiltonian.

To find \(\Delta^z\) and \(\rho^z_\sigma\), we compute
\begin{equation}
    \label{eq:selfconsistent}
    \begin{aligned}
        \Delta^z &= \frac{1}{L_w}\displaystyle\sum_k
        \displaystyle\sum_{n=1}^{2\lmax+1} u_{n,z}^{\phantom\dagger}
        u^*_{n,z+2\lmax+1} \\
        \rho^z_\downarrow &= \frac{1}{L_w}\displaystyle\sum_k
        \displaystyle\sum_{n=1}^{2\lmax+1}
        \abs{u_{n,z+2\lmax+1}^{\phantom\dagger} }^2 \\ 
        \rho^z_\uparrow &= \frac{1}{L_w}\displaystyle\sum_k
        \displaystyle\sum_{n=2\lmax+2}^{2(2\lmax+1)}
        \abs{u_{n,z}^{\phantom\dagger} }^2
    \end{aligned}
\end{equation}
where \(u_{n,m}^{\phantom\dagger} \) is the \(m\)-th element of eigenvector \(u_n\).
Since the eigenvectors are inherently dependent on the mean-field parameters which enter the BdG Hamiltonian, Eq.~\eqref{eq:selfconsistent} are the set of self-consistent equations that have to be satisfied to find the mean-field parameters and ground state of the system.

\section{The Superfluid Weight}
\label{appendix:Ds}

\begin{figure}
    \includegraphics[width=8.6cm]{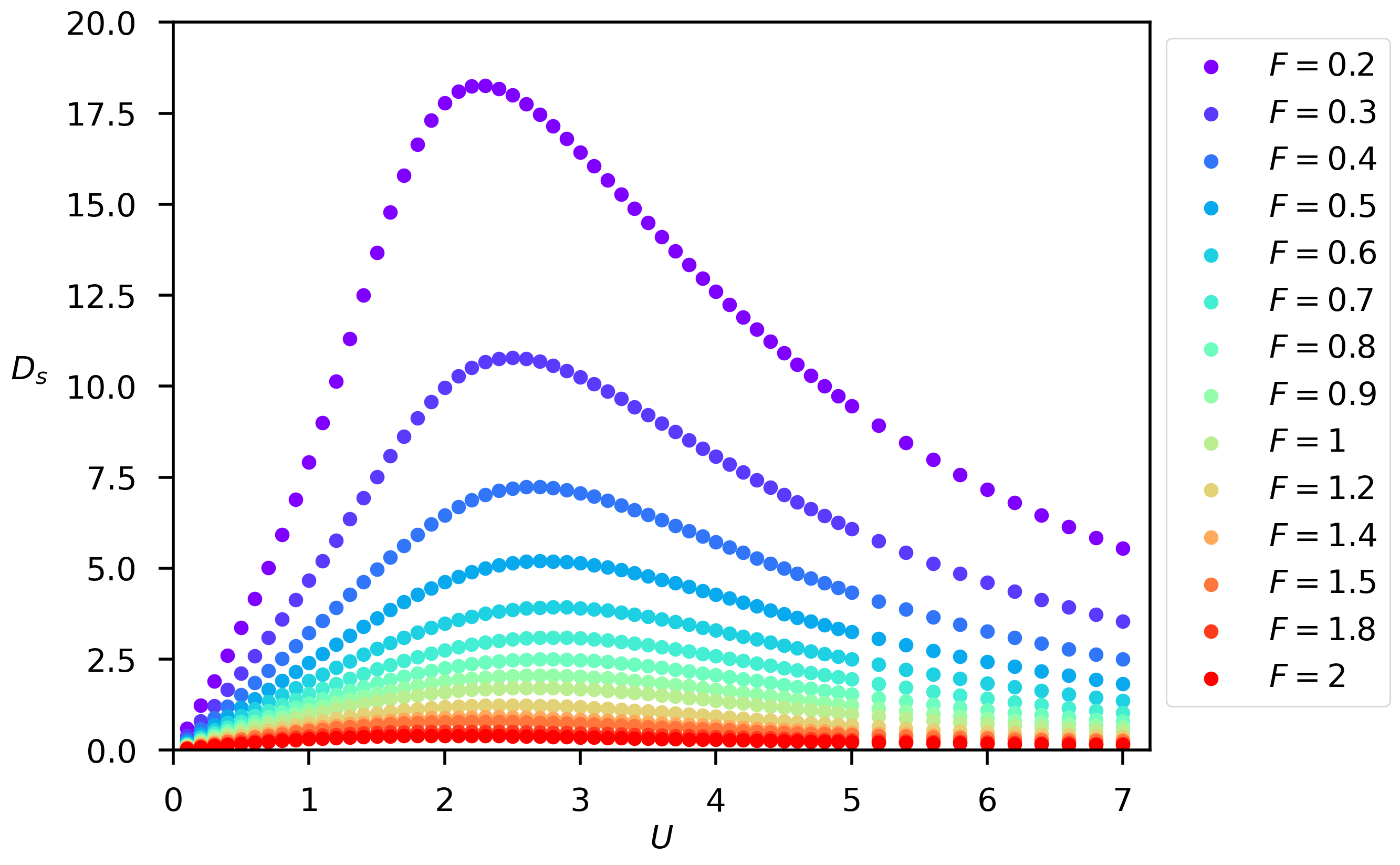}
    \includegraphics[width=8.6cm]{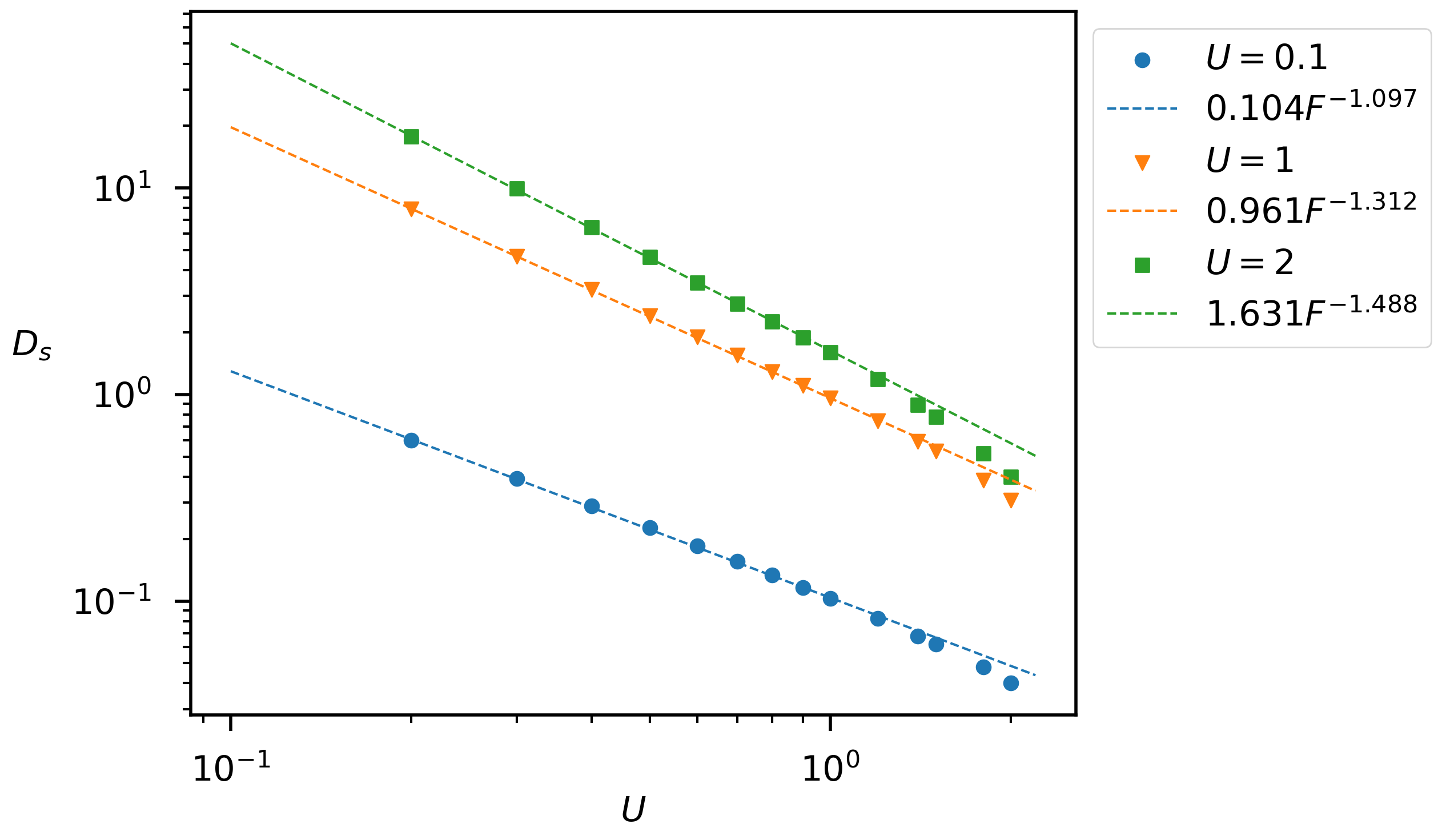}
    \caption{
        (Color online) \textbf{Top}: Superfluid weight on the 2D Wannier-Stark lattice along the \(\mathbf{e_w}\) direction as a function of the attraction strength.
        \textbf{Bottom}: Superfluid weight on the 2D Wannier-Stark flat bands as a function of the DC field strength behaves as a power law at weak interaction.
    }
    \label{fig:Ds2Dcuts}
\end{figure}

To compute the superfluid weight, we determine the ground state energy in the presence of a phase twist, \(\Phi\).

For the 2D square lattice, we apply a phase twist in the \(\mathbf{e_w}\) direction to calculate the superfluid weight.
Specifically, we implement a local gauge transformation on the operators, with \(c_{z,w,\sigma}\rightarrow c_{z,w,\sigma}e^{iw\phi}\) and \(\phi = \frac{\Phi}{\sqrt{2}L_w}\) is the phase gradient.

In the 3D case, we can choose to compute the superfluid weight in any direction, \(\mathbf{e_\delta}\), where \(\delta=n\alpha+m\beta\) on the 2D flat band and replace the operator \(c_{\alpha,\beta,\gamma,\sigma} \rightarrow c_{\alpha,\beta,\gamma,\sigma}e^{i\delta\phi_\delta}\).
The phase gradient, \(\phi_\delta\), depends on the direction we apply the phase twist, for example in the \(\alpha\) direction, the phase gradient is \(\phi_\alpha=\frac{\Phi}{L_\alpha \sqrt{2}a}\), and in the \(\beta\) direction, the phase gradient is \(\phi_\beta=\frac{\Phi}{L_\beta \sqrt{6}a}\).
Following this procedure, the superfluid weight can be shown to be independent of the direction in which we apply the phase twist.

\begin{figure}
    \includegraphics[width=8.6cm]{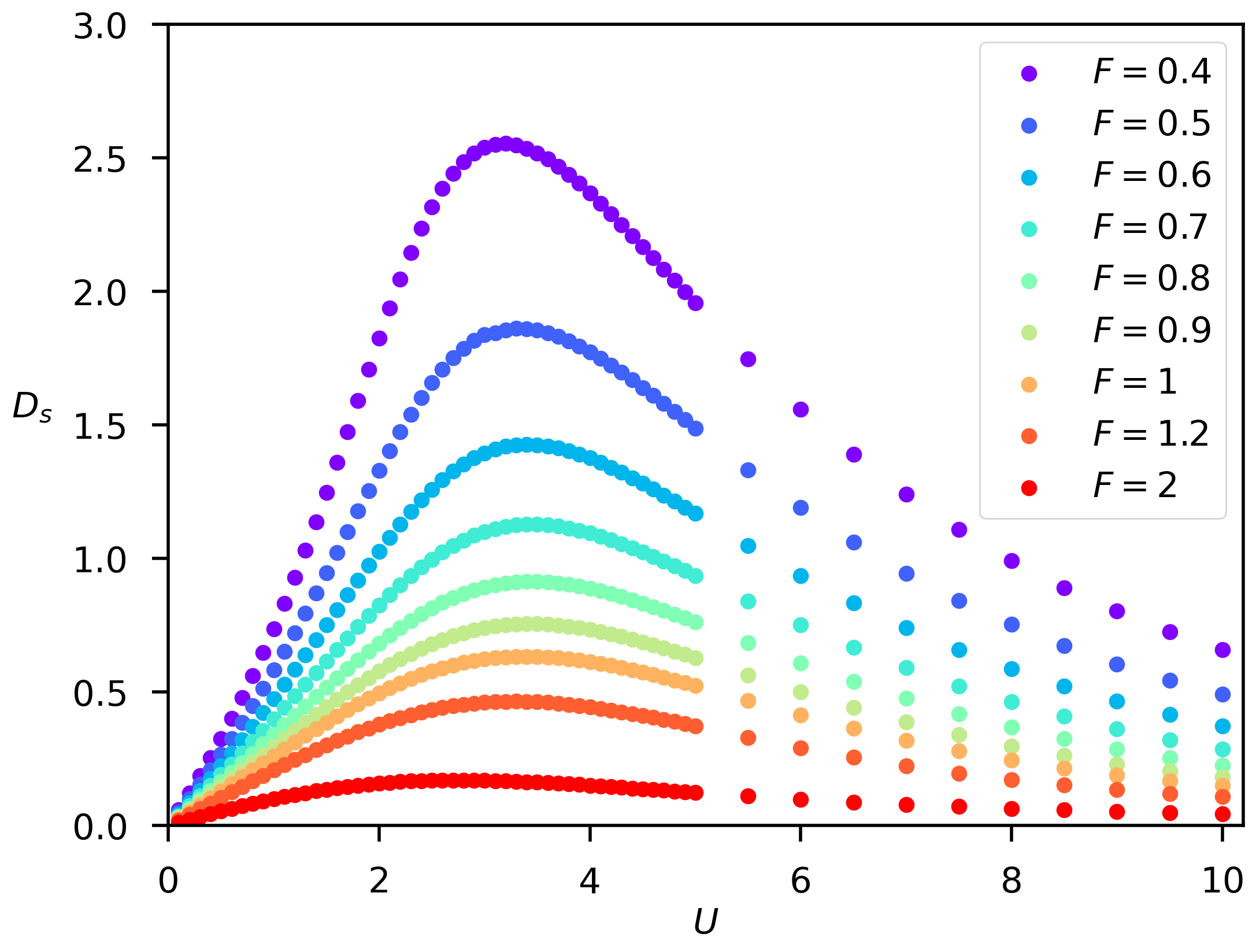}
    \caption{
        (Color online) Superfluid weight perpendicular to \(\mathbf{e_\gamma}\) on the 3D Wannier-Stark lattice as a function of the attraction strength.
    }
    \label{fig:Ds3Dcuts}
\end{figure}

In the main text, we showed how the superfluid weight varies as a function of both \(F\) and \(U\).
Here, we include cuts of the surface plot in Figs.~\ref{fig:Ds2Dcuts} and~\ref{fig:Ds3Dcuts} for the 2D and 3D Wannier-Stark systems respectively.
For any DC field strength, the superfluid weight increases at weak interaction with transport governed by the density overlaps of the single particle wavefunction on the flat bands, peaks at intermediate \(U\) and decreases as the spread of the partially filled flat bands decrease at strong \(U\).

Additionally, we show in Fig.~\ref{fig:Ds2Dcuts} the power law dependence of the superfluid weight on the DC field strength, for \(0.2\leq F\leq 2\), at weak to intermediate interactions of \(U=0.1, 1\) and \(2\).
We see the effect of the wavefunction broadening as \(\frac{t}{F}\) especially evident for \(U=0.1\), where \(D_s \sim F^{-1.097}\).

\bibliography{WannierStarkHubbard}

\end{document}